\documentclass[12pt, draftclsnofoot, onecolumn]{IEEEtran}
\usepackage[caption=false,font=footnotesize]{subfig}
\usepackage{graphicx}
\usepackage{booktabs}
\usepackage{amsmath}
\usepackage{amssymb}
\usepackage{amsthm}
\usepackage{midfloat}
\usepackage{bm} 
\usepackage{color, cite}
\usepackage{algorithm}
\graphicspath{{figures/}}

\newtheorem{lemma}{Lemma}

\begin{document}

\title{Multi-target Position and Velocity Estimation Using OFDM Communication Signals}

\author{
Yinchuan~Li, Xiaodong Wang, \emph{Fellow}, \emph{IEEE}, Zegang~Ding, \emph{Member}, \emph{IEEE}% <-this % stops a space
%\thanks{This work was supported by the Key Program of National Natural Science Foundation of China (No. 11833001), the National Science Fund for Distinguished Yong Scholars (No. 61625103) and the China Scholarship Council.}%
\thanks{Y. Li, and Z. Ding are with the School of Information and Electronics, Beijing Institute of Technology, Beijing 100081, China, and the Beijing Key Laboratory of Embedded Real-time Information Processing Technology, Beijing 100081, China (e-mail: yinchuan.li.cn@gmail.com; z.ding@bit.edu.cn).

Y. Li and X. Wang are with the Electrical Engineering Department, Columbia University, New York, NY 10027, USA (e-mail: wangx@ee.columbia.edu).}

}%

\maketitle

\begin{abstract}

In this paper, we consider a passive radar system that estimates the positions and velocities of multiple moving targets by using OFDM signals transmitted by a totally un-coordinated and un-synchronizated illuminator and multiple receivers. It is assumed that data demodulation is performed separately based on the direct-path signal, and the error-prone estimated data symbols are made available to the passive radar receivers, which estimate the positions and velocities of the targets in two stages. First, we formulate a problem of joint estimation of the delay-Doppler of reflectors and the demodulation errors, by exploiting two types of sparsities of the system, namely, the numbers of reflectors (i.e., targets and clutters) and demodulation errors are both small. This problem is non-convex and a conjugate gradient descent method is proposed to solve it. Then in the second stage we determine the positions and velocities of targets based on the estimated delay-Doppler in the first stage. And two methods are proposed: the first is based on numerically solving a set of nonlinear equations, while the second is based on the back propagation neural network, which is more efficient. The performance of the proposed passive OFDM radar receiver algorithm is evaluated through extensive simulations.

% by a communication receiver It is seen that accurate target estimation results are obtained even with high demodulation error rate.
\end{abstract}

\section{Introduction}

Passive radar systems can detect targets by utilizing readily available, non-cooperative illuminators of opportunity (IOs)~\cite{liu2014two,fang2018experimental,zheng2017super,hack2014centralized}, and possess a number of benefits compared with active radar systems. In particular, a passive radar is smaller and less expensive because it does not need a transmitter. And many IOs (e.g., cellular base stations~\cite{sun2010applications}, analog television broadcasting~\cite{howland1999target}, digital audio broadcasting (DAB)~\cite{poullin2005passive}) are available for passive sensing, as such a passive radar can operate without causing interference to existing communication systems.

A main challenge associated with passive radar is that the IOs are non-cooperative, and the transmitted signals are unknown and not under control. Hence, the conventional matched filter cannot be easily implemented. In addition, the direct-path signal is much stronger than the target reflections, making it difficult to detect and track targets. To solve those problems, a passive radar usually makes use of an additional separate channel, referred to as the reference channel, to collect the transmitted signal in order to eliminate the direct-path signal and clutters in the surveillance channels (SCs)~\cite{tao2010direct,cardinali2007comparison}. In addition, the reference signal can also be used to implement approximate matched filtering to detect targets. However, the reference signal is noisy and the target detection performance is usually significantly degraded~\cite{liu2014two,zheng2017super}.

Orthogonal frequency-division multiplexing (OFDM) techniques are widely employed in many modern wireless communication systems, e.g., 4G wireless cellular~\cite{salah2014experimental}, digital video broadcasting (DVB)~\cite{palmer2013dvb}, DAB~\cite{poullin2005passive} and wireless local area network (LAN)~\cite{falcone2010experimental,colone2011ambiguity}. For an OFDM passive radar system, demodulation can be implemented by using the reference signal~\cite{colone2009multistage,berger2010signal,zheng2017super}. Since demodulation provides better accuracy than directly using the reference signal, a more accurate matched filter can be implemented based on the estimated data symbols, and the performance of passive radar can be greatly improved. Moreover, the reference channel is not always necessary because data symbols can also be directly demodulated based on the received signal in SC.

Some target detection algorithms using OFDM signals have been proposed~\cite{zheng2017super,palmer2013dvb,berger2010signal,sen2011adaptive,falcone2010experimental}. In~\cite{sen2011adaptive}, a method for detecting a moving target in the presence of multi-path reflections is proposed based on adaptive OFDM radar. In~\cite{berger2010signal,palmer2013dvb,falcone2010experimental}, by assuming that the demodulation is perfect, the delays and Doppler shifts of targets are estimated based on matched filtering. And in~\cite{berger2010signal}, the MUltiple SIgnal Classifier (MUSIC) and the compressed sensing (CS) techniques are employed to obtain a better target resolution~\cite{ding2016modified} and clutter removal performance. In~\cite{zheng2017super}, by using the received OFDM signal from an un-coordinated but synchronizated illuminator, a delay and Doppler shift estimation algorithm is proposed taking into account the demodulation error. The atomic norm (AN) is used to enforce the signal sparsity in the delay-Doppler plane and the $\ell_1$-norm is used to enforce the sparsity of the demodulation error signal. Then, a convex semidefinite program (SDP) is solved to obtain the estimate of the target delays and Doppler shifts.

The present contribution is aimed at extending the results of~\cite{zheng2017super} by using multiple receivers to achieve the target position and velocity estimation based on the OFDM signal emitted by a totally un-coordinated and un-synchronizated illuminator. Assuming that data demodulation is performed separately by a communication receiver based on the direct-path signal, and the error-prone estimated data symbols are made available to the passive radar receivers. Then, a two-stage procedure is proposed to estimate the positions and velocities of targets.

The first stage is aimed at estimating the delay-Doppler shift and demodulation error by exploiting two types of sparsity: on one hand, as targets and clutters are sparsely distributed in space, the reflected signals hitting the radar receivers are sparse; on the other, the demodulation error rate of a communication system is typically low under normal operating conditions and hence the demodulation error signal is also sparse. Since the delays and Doppler shifts of the targets are continuous parameters, conventional CS tools~\cite{candes2011compressed} may lead to unsatisfactory performance~\cite{chi2011sensitivity} when the signals cannot be sparsely represented by a finite discrete dictionary~\cite{stankovic2013compressive,jokanovic2015reduced,studer2012recovery}. We make use of the recently developed mathematical theory of continuous sparse recovery for super-resolution~\cite{candes2013super,candes2014towards,tang2013compressed}, and especially the AN minimization techniques which have been successfully applied for continuous frequency recovery, line spectral estimation and direction-of-arrival estimation~\cite{tang2013compressed,bhaskar2013atomic,tan2014direction,li2019interference}. Note that, unlike the convex problem of the delay-Doppler estimation in~\cite{zheng2017super} for one receiver, in our model, different receivers share the same estimated data symbols and impose the same constraint, which yields a non-convex problem due to existence of the produce term of decision variables. Hence, we use non-convex factorization (NF) to transform the problem to a smooth unconstrained optimization problem, which is then solved by a conjugate gradient descent (CGD) algorithm.

The second stage is aimed at determining the target positions and velocities based on the estimates in the first stage. Since the illuminator and receivers are un-synchronizated, we utilize the delay differences between different receivers to calculate each target position. The first method numerically solves a set of nonlinear equations, and the second method utilizes the back propagation (BP) neural network~\cite{liang2016lagrange,yang2018fast,jia2015bp} to estimate the target position, which is more computationally efficient. The corresponding target velocity can then be determined based on the estimated target position and Doppler shift. Extensive simulation results are provided to illustrate that the proposed methodology can estimate the target positions and velocities accurately.

The remainder of the paper is organized as follows. In Section II, we present the signal model of the OFDM passive radar and set up the problem. In Section III, we develop a delay-Doppler estimator based on conjugate gradient descent. In Section IV, we discuss methods for estimating the locations and velocities. Simulation results are presented in Section V. Section VI concludes the paper.

\section{System Descriptions \& Problem Formulation}

\subsection{System Descriptions}

\begin{figure}%[!htb]
	\centering
		
	\subfloat{\includegraphics[width=3.0in]{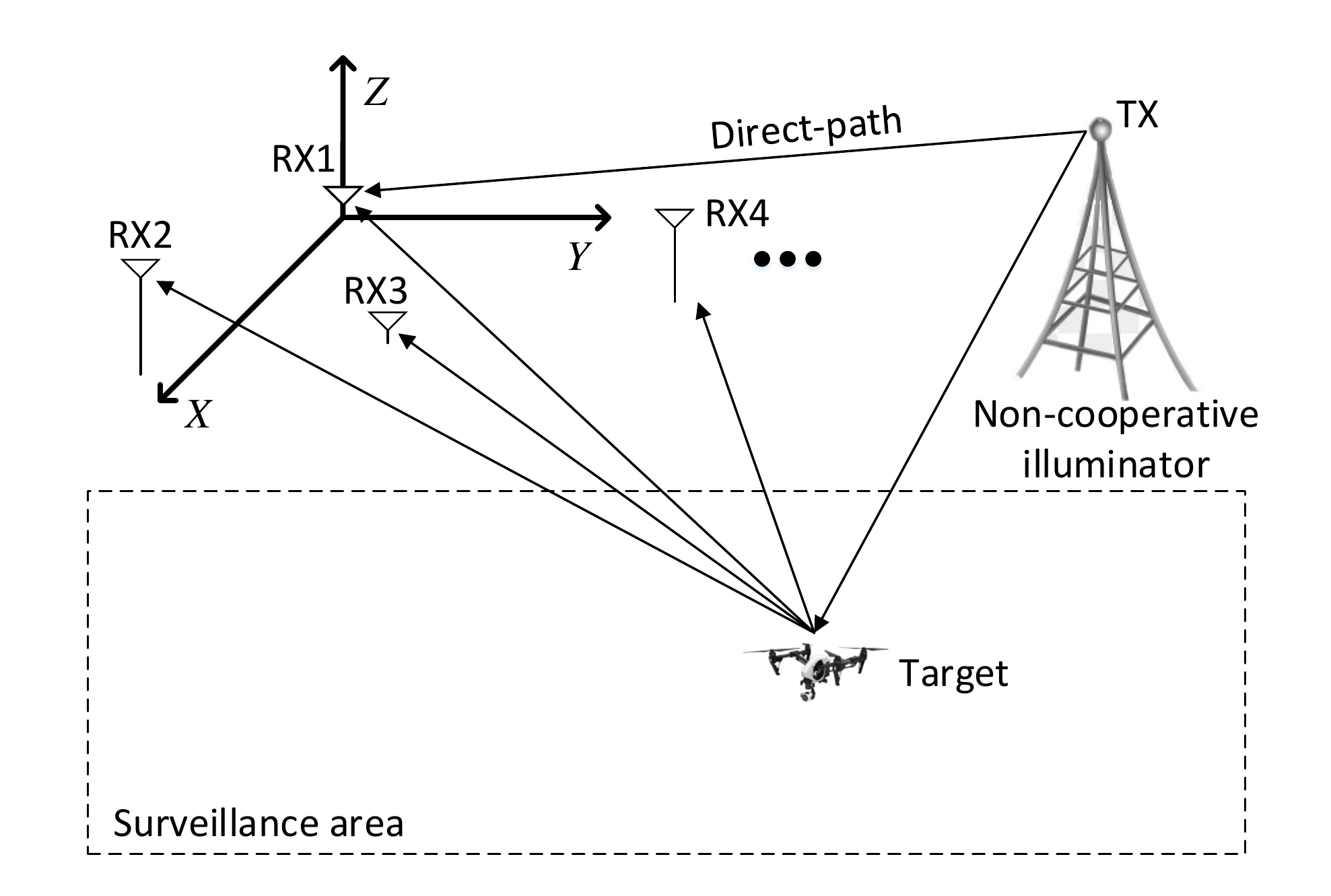}}
	
	\caption{A passive radar system for target location and velocity estimation.}
	\label{figure:drone}
\end{figure}

As shown in Fig.~\ref{figure:drone}, we consider a passive radar system consisting of $M$ ($M\geq4$) receivers and one non-cooperative illuminator, that aims to estimate the locations and velocities of multiple targets in a three-dimensional cartesian coordinate system. Suppose that the coordinates of the illuminator and receiver $m$ are $\bm p_0 = [p^x_0,p^y_0,p^z_0]^T$ and $\bm p_m = [p^x_m,p^y_m,p^z_m]^T,$ $m=1,...,M$, respectively. Assume that there are $L$ reflectors in the surveillance area, include targets and clutters. Note that we consider clutters as zero-velocity targets. Let $\bm x_{\ell} = [x^x_{\ell},x^y_{\ell},x^z_{\ell}]^T$ and $\bm v_{\ell} = [v_{\ell}^x,v_{\ell}^y,v_{\ell}^z]^T$ be the location and velocity of the $\ell$-th reflector, respectively. Then, the traveling time from the illuminator to the $m$-th receiving antenna due to the $\ell$-th reflector is
\begin{eqnarray}
\label{eq:tau}
{\bar\tau}_{\ell,m} = \frac{1}{c}(\| \bm p_0 - \bm x_{\ell}\|_2 + \| \bm p_m - \bm x_{\ell}\|_2),
\end{eqnarray}
where $c$ is the speed of light in free-space; $\|\cdot\|_2$ denotes the $\ell_2$-norm. And the corresponding Doppler shift is given by~\cite{he2010noncoherent}
\begin{align}
\label{eq:doppler}
{\bar f}_{\ell,m} =&~ \frac{\bm v_{\ell}^T(\bm p_0 - \bm x_{\ell})}{\lambda \| \bm p_0 - \bm x_{\ell}\|_2} + \frac{\bm v_{\ell}^T(\bm p_m - \bm x_{\ell})}{\lambda \| \bm p_m - \bm x_{\ell}\|_2},
\end{align}
where $\lambda$ denotes the wavelength of the carrier.

Assume that $s(t)$ is the unknown communication signal. Due to the reflections of targets and clutters, the received signal at the $m$-th receiver is given by
\begin{align}
\label{eq:barst}
y_m(t) =&~ s^d_m(t) + \sum_{\ell=1}^{L} c_{\ell,m} e^{i2\pi {\bar f}_{\ell,m} t} s(t - {\bar\tau}_{\ell,m} + \Delta_{\tau}) + {w}_m(t),
\end{align}
where $c_{\ell,m}$ is the $\ell$-th path's complex gain at the $m$-th receiving antenna{\footnote{In this paper, we assume that the complex gains do not vary from pulse to pulse, e.g., Swerling reflectors of types 0 and 1.}}; $s^d_m(t)$ is the direct-path (illuminator-to-receiver) signal{\footnote{Note that the direct-path signal $s^d_m(t)$ can be suppressed by the spatial filtering method in~\cite{tao2010direct} or by using a reference channel to collect the direct-path signal~\cite{liu2014two,fang2018experimental,liu2015performance}, i.e., using a narrow beam antenna towards the transmitter to receive the direct-path signal~\cite{liu2015performance}, or using the side-lobe of the receiving antenna to receive the direct-path signal~\cite{fang2018experimental}.}; ${w}_m(t)$ is a white, complex circularly symmetric Gaussian process; and $\Delta_{\tau}$ is the synchronization error between the transmitter and receivers, i.e., we assume that the radar receivers share the same clock but are not synchronized according to the communication transmitter.
%$\ast$ denotes the convolution operator; $\delta(\cdot)$ is the unit impulse function;

\subsection{OFDM-based Passive Radar Signal Model}
In this paper, we assume that the signal $s(t)$ is the OFDM signal that is widely adopted in contemporary wireless communication systems. The OFDM system consists of $N_d$ data sub-carriers and $NT = (N_d + N_p)T$ basic time units, with $N_p$ being the number of cyclic prefix (CP) carriers and $T$ being the sampling period (``sub-pulse duration''). 
%$N = N_d+N_p$ sub-carriers, with $N_d$ data sub-carriers and $N_p$ cyclic prefix (CP) sub-carriers. The duration of an OFDM symbol is $NT$ with $T$ being the ``sub-pulse duration.'' 
Then, the transmitted baseband OFDM signal over $N_b$ blocks is given by
\begin{equation}
\label{eq:st}
s(t)= \sum_{n=0}^{N_b-1} \sum_{k=0}^{N_d-1} b_n(k)e^{i2\pi k \frac{t}{N_dT} } u(t - n NT),
\end{equation}
where $b_n(k), ~k = 0,...,N_d-1$ is the $n$-th normalized data symbol block, such that $\mathbb{E}[b_n(k)b_n(k)^*]=1$ with $(\cdot)^*$ denoting the complex conjugate operator; and $u(t) = \left\{
   \begin{aligned}
   1,&~~t\in[-N_pT,N_dT],  \\
   0,&~~\text{otherwise}. \\
   \end{aligned}
   \right.$
%\begin{equation}
%\label{eq:ut}
%u(t) = \left\{
%   \begin{aligned}
%   1,&~~t\in[-N_pT,N_dT],  \\
%   0,&~~\text{otherwise}. \\
%   \end{aligned}
%   \right.
%\end{equation}

At each radar receiver $m$, suppose that the direct-path signal $y_m^d(t)$ is first removed, and we only refer to the baseband signals by assuming that down-conversion has been performed. The CP is removed assuming that its length is no less than the maximum path delay, i.e., $N_pT> \max_{m,\ell}\{{\bar\tau}_{\ell,m} - \Delta_{\tau}\}$. Note that the data symbols $b_n(k)$ are unknown, but can be estimated by demodulation using the direct-path signals $s^d_m(t)$~\cite{liu2014two}. However, demodulation may be error-prone{\footnote{We assume that the passive radar system only performs demodulation of the data symbols, but not the forward error correction (FEC) decoding~\cite{ji2005rate}. Since the code book may not be available to the passive radar system for security or privacy reasons, and the FEC also increases the complexity of the radar signal processing.}}. Hence in the following we assume that an estimate of the data symbols, $\hat b_n(k)$, is available such that 
\begin{align}
b_n(k) = \hat b_n(k) + e_n(k),~k = 0,...,N_d-1,
\end{align}
where $e_n(k)$ denotes the corresponding demodulation error. Furthermore, we assume that the velocity of the target is low, such that ${\bar f}_{\ell,m}NT \ll 1$. Hence the phase rotation due to the Doppler shift can be approximated as constant over an OFDM symbol duration $NT$, i.e.,~\cite{berger2010signal,zheng2017super}
\begin{align}
e^{i2\pi {\bar f}_{\ell,m} t} \approx e^{i2\pi {\bar f}_{\ell,m} n {NT}}, ~ t \in [nNT,(n+1)NT].
\end{align}
At each receiver $m$, in the $n$-th OFDM symbol matched filtering is performed to obtain, for $k = 0,...,N_d-1$,
\begin{align}
&\bar y_{n,m}(k) = \frac{1}{N_dT} \int_{nNT}^{nNT+N_d T} {(y_{m}(t)-s^d_m(t)) {e^{\frac{{ - i2\pi k t}}{N_dT}}}} dt + \bar w_{n,m}(k)\nonumber \\
=&~ { \sum_{\ell=1}^{L} c_{\ell,m} } { \sum_{q=0}^{N_d-1} b_n(q)  \frac{1}{N_dT}  \int_{nNT}^{nNT+N_d T} \underbrace{e^{i2\pi {\bar f}_{\ell,m} t}}_{\approx e^{i2\pi {\bar f}_{\ell,m} n {NT}}} e^{i2\pi q \frac{t - {\bar\tau}_{\ell,m} + \Delta_{\tau} }{N_dT} } {e^{\frac{{ - i2\pi k t}}{N_dT}}}} dt + \bar w_{n,m}(k) \nonumber \\
\label{eq:ynm1}
\approx&~ { \sum_{\ell=1}^{L} c_{\ell,m} } e^{i2\pi {\bar f}_{\ell,m} n {NT}}  { \sum_{q=0}^{N_d-1} b_n(q)  e^{-i2\pi q \frac{{\bar\tau}_{\ell,m} - \Delta_{\tau}}{N_dT}}  \frac{1}{N_dT} \underbrace{ \int_{nNT}^{nNT+N_d T} e^{i2\pi (q-k) \frac{t}{N_dT}} dt }_{N_dT \cdot \delta(q-k) } } + \bar w_{n,m}(k) \\
\label{eq:ynm2}
=&~ (\hat b_n(k) + e_n(k))    \sum\limits_{\ell = 1}^{L} c_{\ell,m} e^{i2\pi n f_{\ell,m}} {e^{ - i2\pi k \tau_{\ell,m}}} + \bar w_{n,m}(k),
\end{align}
where $\bar w_{n,m}(k) = \frac{1}{N_dT} \int_{nNT}^{nNT+N_d T} {{w}_m(t) {e^{\frac{{ - i2\pi k t}}{N_dT}}}} dt$ and
\begin{align}
\label{eq:tau-DeltaM}
\tau_{\ell,m} = \frac{{\bar\tau}_{\ell,m} - \Delta_{\tau}}{N_dT} \in [0,1),~f_{\ell,m} =&~ {\bar f}_{\ell,m}{NT} \in [0,1).
\end{align}

\subsection{Problem Formulation}

Let us now define $\bm c_m = [c_{1,m},c_{2,m},...,c_{L,m}]^T \in \mathbb{C}^{L \times 1}$, $\bm f_m = [f_{1,m},f_{2,m},...,f_{L,m}]^T \in \mathbb{C}^{L \times 1}$ and $\bm \tau_m = [\tau_{1,m},\tau_{2,m},...,\tau_{L,m}]^T \in \mathbb{C}^{L \times 1}$,
%\begin{align}
%\bm c_m =&~ [c_{1,m},c_{2,m},...,c_{L,m}]^T \in \mathbb{C}^{L \times 1}, \\
%\bm f_m =&~ [f_{1,m},f_{2,m},...,f_{L,m}]^T \in \mathbb{C}^{L \times 1}, \\
%\bm \tau_m =&~ [\tau_{1,m},\tau_{2,m},...,\tau_{L,m}]^T \in \mathbb{C}^{L \times 1},
%\end{align}
and the steering vectors $\bm s(f) = [1,e^{i2\pi f},...,e^{i2\pi(N_b-1)f}]^T \in \mathbb{C}^{N_b \times 1}$ and $\bm d(\tau) = [1,e^{i2\pi\tau},...,e^{i2\pi(N_d-1)\tau}]^T \in \mathbb{C}^{N_d \times 1}$.
%\begin{align}
%\bm s(f) =&~ [1,e^{i2\pi f},...,e^{i2\pi(N_b-1)f}]^T \in \mathbb{C}^{N_b \times 1}, \\
%\bm d(\tau) =&~ [1,e^{i2\pi\tau},...,e^{i2\pi(N_d-1)\tau}]^T \in \mathbb{C}^{N_d \times 1}.
%\end{align}
Correspondingly, the response matrices are defined as $\bm S(\bm f_m) = [\bm s(f_{1,m}), \bm s(f_{2,m}), ..., \bm s(f_{L,m})] \in \mathbb{C}^{N_b \times L}$ and $\bm D(\bm\tau_m) = [\bm d(\tau_{1,m}), \bm d(\tau_{2,m}), ..., \bm d(\tau_{L,m})] \in \mathbb{C}^{N_d\times L}$.
%\begin{align}
%\bm S(\bm f_m) =&~ [\bm s(f_{1,m}), \bm s(f_{2,m}), ..., \bm s(f_{L,m})] \in \mathbb{C}^{N_b \times L}, \\
%\bm D(\bm\tau_m) =&~ [\bm d(\tau_{1,m}), \bm d(\tau_{2,m}), ..., \bm d(\tau_{L,m})] \in \mathbb{C}^{N_d\times L}.
%\end{align}
Then \eqref{eq:ynm2} can be written as the following matrix form
\begin{eqnarray}
\label{eq:Y2}
\bm{\bar Y}_m = (\bm{\hat{B}} + \bm E) \odot (\bm S(\bm f_m) {\rm{diag}}(\bm c_m) \bm D(\bm\tau_m)^H) + \bm{\bar W}_m,
\end{eqnarray}
where $\odot$ denotes the Hadamard product; ${\rm{diag}}(\bm c_m)$ denotes the diagonal matrix whose diagonal entries are $\bm c_m$; $\bm{\bar Y}_m \in \mathbb{C}^{N_b\times N_d}$, $\bm{\hat{B}} \in \mathbb{C}^{N_b\times N_d}$, $\bm E \in \mathbb{C}^{N_b \times N_d}$ and $\bm{\bar W}_m \in \mathbb{C}^{N_b\times N_d}$ are matrices whose $(n,k)$-th element are $\bar y_{n,m}(k)$, $\hat b_n(k)$, $e_n(k)$ and $\bar w_{n,m}(k)$, respectively.

Further denote  $\bm{\hat b} = {\rm{vec}}(\bm{\hat{B}}) \in \mathbb{C}^{N_bN_d\times 1}$, $\bm e = {\rm{vec}}(\bm E) \in \mathbb{C}^{N_bN_d\times 1}$, $\bm{\bar w}_m = {\rm{vec}}(\bm{\bar W}_m) \in \mathbb{C}^{N_bN_d\times 1}$ and
\begin{align}
\label{eq:nu}
\bm\phi_m =&~ \sum_{\ell=1}^{L}c_{\ell,m} \bm a(\tau_{\ell,m},f_{\ell,m}) \in \mathbb{C}^{N_bN_d\times 1}, \\
\label{eq:atau-f}
\text{with}~\bm a(\tau,f) =&~ \bm d(\tau)^*\otimes \bm s(f) \in \mathbb{C}^{N_bN_d\times 1},
\end{align}
and $\otimes$ being the Kronecker product. Then we vectorize $\bm{\bar Y}_m$ in \eqref{eq:Y2} to obtain
\begin{align}
\label{eq:y1}
\bm{\bar y}_m =&~ {\rm{vec}}(\bm{\bar Y}_m) ={\rm{diag}}(\bm{\hat b} + \bm e) \left( \bm D(\bm\tau_m)^* \circ \bm S(\bm f_m) \right) \bm c_m  + \bm{\bar w}_m = {\rm{diag}}(\bm{\hat b} + \bm e) \bm\phi_m  + \bm{\bar w}_m,
\end{align}
where $\circ$ is the Khatri-Rao product;  $\left( \bm D(\bm\tau_m)^* \circ \bm S(\bm f_m)\right) \in \mathbb{C}^{N_bN_d\times L}$ is a matrix whose $\ell$-th column has the form of $\bm d^*(\tau_{\ell,m})  \otimes \bm s(f_{\ell,m})$. Finally, \eqref{eq:y1} can be rewritten in the following matrix form
\begin{eqnarray}
\label{eq:problem}
\bm Y = {\rm{diag}}(\bm{\hat b} + \bm e) \bm\Phi  + \bm W,
\end{eqnarray}
where the $m$-th columns of $\bm Y\in \mathbb{C}^{N_bN_d\times M}$, $\bm\Phi\in \mathbb{C}^{N_bN_d\times M}$ and $\bm W\in \mathbb{C}^{N_bN_d\times M}$ are $\bm{\bar y}_m$, $\bm\phi_m$ and $\bm{\bar w}_m$, respectively.

In this paper, we first estimate the delays and Doppler shifts $\{\tau_{\ell,m},f_{\ell,m}\}$ contained in $\bm \Phi$ from the received signals $\bm Y$. Then based on these estimates, we further estimate the locations and velocities of the reflectors $\{\bm x_{\ell},\bm v_{\ell}\}$, and those with $\bm v_{\ell}\approx 0$ are considered clutters.

\section{Stage 1: Delay-Doppler Estimation}

In this section, we propose a CGD method to estimate the delays and Doppler shifts $\{\tau_{\ell,m},f_{\ell,m}\}$ in \eqref{eq:problem}. We first formulate a non-convex optimization problem by exploiting two types of sparsity. Then we relax the non-convex optimization problem to a smooth unconstrained form. The smoothed problem can then be solved via CGD.

\subsection{Non-convex Optimization Problem Setup}

We will exploit the following two types of sparsity: firstly, the number of reflectors $L\ll N_bN_d$ in \eqref{eq:nu}; secondly, assuming that the demodulation error rate is low, then $\bm e$ has a small number of non-zero entries, i.e., $\|\bm e\|_0\ll N_bN_d$ with $\|\cdot\|_0$ being the $\ell_0$-norm. Since the delays $\bm\tau$ and the Doppler shifts $\bm f$ take continuous values, the atomic norm~\cite{tang2013compressed,yang2016super} is used to exploit the first type of sparsity. Let ${\cal A} = \{ \bm a(\tau,f): \tau\in[0,1), f\in[0,1)\}$ be the set of atoms, where $\bm a(\tau,f)$ is defined in \eqref{eq:atau-f}. Then the 2D atomic norm~\cite{tang2013compressed} associated to $\bm\phi_m$ for $m=1,...,M$ is defined as
\begin{eqnarray}
\label{eq:2D-AN}
\| \bm\phi_m \|_{{\cal A}} &=& \inf \left\{ \chi>0: \bm\phi_m \in \chi {\rm conv}({\cal A}) \right\} \nonumber \\
&=& \inf_{\substack{ c_{\ell,m}\in{\mathbb{C}}, f_{\ell,m}\in[0,1),\\ \tau_{\ell,m} \in[0,1) } } \left\{ \sum_{\ell} |c_{\ell,m}|: \bm\phi_m = \sum_{\ell} c_{\ell,m} \bm a(\tau_{\ell,m},f_{\ell,m}) \right\}.
\end{eqnarray}
The atomic norm can enforce sparsity in the atom set ${\cal A}$. Note that columns in $\bm\Phi$ are independent with their own sparsities. On this basis, our delay-Doppler estimation problem can be formulated according to \eqref{eq:problem} as:
\begin{eqnarray}
\label{eq:optimization-problem}
(\bm{\hat\Phi}, \bm{\hat e}) = \arg \mathop {\min }\limits_{\substack{ \bm\Phi \in \mathbb{C}^{N_bN_d \times M},\\ \bm e \in \mathbb{C}^{N_bN_d \times 1} } }  \frac{1}{2}\| \bm Y - {\rm{diag}}(\bm{\hat b} + \bm e) \bm\Phi \|_F^2  + \gamma \sum_{m=1}^{M}\| \bm\phi_m \|_{{\cal A}} + \eta \| \bm e \|_1,
\end{eqnarray}
where $\|\cdot\|_1$ denotes the $\ell_1$-norm, $\gamma>0$ and $\eta>0$ are the weight factors.

However, finding the harmonic components via atomic norm is an infinite programming problem over all feasible $\bm \tau$ and $\bm f$. For the convenience of calculation, we use the following equivalent form of \eqref{eq:2D-AN} for $m=1,...,M$~\cite{tang2013compressed,chi2015compressive}
\begin{eqnarray}
\label{eq:atomic}
\| \bm{\phi}_m \|_{\cal {A}} = \mathop {\inf}\limits_{\substack{ \bm Q_m \in \mathbb{C}^{(2N_b-1) \times (2N_d-1)}, \\ \nu_m \in \mathbb{R} } } \left\{ \begin{array}{l}
\frac{1}{2N_bN_d}{\rm{Tr}}({\mathbb{T}}(\bm Q_m)) + \frac{\nu_m}{2},\\
{\rm s.t.} \left[ {\begin{array}{*{20}{c}}
	{{\mathbb{T}}(\bm Q_m)}& \bm{\phi}_m\\
	{\bm{\phi}_m^H}& {\nu_m}
	\end{array}} \right] \succeq 0
\end{array} \right\},
\end{eqnarray}
where ${\rm{Tr}}(\cdot)$ denotes the trace operator, $\succeq 0$ stands for a positive semidefinite matrix, and ${\mathbb{T}}(\cdot)$ takes as input a ${(2N_b-1)\times (2N_d-1)}$ matrix 
\begin{eqnarray}
\bm Q_m = [\bm q_{m,-N_d+1},\bm q_{m,-N_d+2},...,\bm q_{m,N_d-1}] \in \mathbb{C}^{(2N_b-1)\times (2N_d-1)},
\end{eqnarray}
with 
\begin{eqnarray}
\bm q_{m,n} = [q_{m,n}(-N_b+1),q_{m,n}(-N_b+2),...,q_{m,n}(N_b-1)]^T \in \mathbb{C}^{(2N_b-1)\times 1}, \\
n = -N_d+1,-N_d+2,...,N_d-1, \nonumber
\end{eqnarray}
and outputs an ${N_bN_d \times N_bN_d}$ block Toeplitz matrix
\begin{eqnarray}
\label{eq:block-Toeplitz}
{\mathbb{T}}(\bm Q_m) = \left[ {\begin{array}{*{20}{c}}
	{{\rm Toep}(\bm q_{m,0})} & {{\rm Toep}(\bm q_{m,-1})} & \cdots & {{\rm Toep}(\bm q_{m,-N_d+1})}\\
	{{\rm Toep}(\bm q_{m,1})} & {{\rm Toep}(\bm q_{m,0})} & \cdots & {{\rm Toep}(\bm q_{m,-N_d+2})}\\
	\vdots & \vdots & \ddots & \vdots \\
	{{\rm Toep}(\bm q_{m,N_d-1})} & {{\rm Toep}(\bm q_{m,N_d-2})} & \cdots & {{\rm Toep}(\bm q_{m,0})}
	\end{array}} \right] \in \mathbb{C}^{N_bN_d \times N_bN_d},
\end{eqnarray}
where ${\rm Toep}(\cdot)$ denotes the Toeplitz matrix whose first column is the last $N_b$ elements of the input vector. More specifically, we have
\begin{eqnarray}
{{\rm Toep}(\bm q_{m,n})} = \left[ {\begin{array}{*{20}{c}}
	{q_{m,n}(0)} & {q_{m,n}(-1)} & \cdots & {q_{m,n}(-N_b+1)}\\
	{q_{m,n}(1)} & {q_{m,n}(0)} & \cdots & {q_{m,n}(-N_b+2)}\\
	\vdots & \vdots & \ddots & \vdots \\
	{q_{m,n}(N_b-1)} & {q_{m,n}(N_b-2)} & \cdots & {q_{m,n}(0)}
	\end{array}} \right] \in \mathbb{C}^{N_b \times N_b}, \\
	n = -N_d+1,-N_d+2,...,N_d-1. \nonumber
\end{eqnarray}
Equations \eqref{eq:2D-AN} and \eqref{eq:atomic} are related when achieving the optimum through the relationship 
\begin{align}
{\mathbb{T}}(\bm Q_m)  =&~ \sum_{\ell,m} |c_{\ell,m}| \bm a(\tau_{\ell,m},f_{\ell,m}) \bm a(\tau_{\ell,m},f_{\ell,m})^H, \\
\nu_m =&~ \sum_{\ell,m} |c_{\ell,m}|.
\end{align}

By using \eqref{eq:atomic}, \eqref{eq:optimization-problem} can be transformed to the following optimization problem:
\begin{align}
\label{eq:SDP1}
(\bm{\hat\Phi}, \bm{\hat e}) =&~ \arg \mathop {\min }\limits_{\substack{ \bm\Phi \in \mathbb{C}^{N_bN_d \times M},\\ \bm Q_m \in \mathbb{C}^{(2N_b-1) \times (2N_d-1)}, \\ \bm e \in \mathbb{C}^{N_bN_d \times 1} } }  \frac{1}{2}\| \bm Y - {\rm{diag}}(\bm{\hat b} + \bm e) \bm\Phi \|_F^2  + \frac{\gamma}{2N_bN_d} \sum\limits_{m=1}^{M} {\rm{Tr}}({\mathbb{T}}(\bm Q_m)) +\frac{\gamma}{2}\sum\limits_{m=1}^{M} \nu_m + \eta \| \bm e \|_1, \nonumber \\
 &~{\rm s.t.}~\left[ {\begin{array}{*{20}{c}}
	{\mathbb{T}}(\bm Q_m)& \bm{\phi}_m\\
	{\bm{\phi}_m^H}& {\nu_m}
	\end{array}} \right] \succeq 0,~m=1,...,M.
\end{align}
Note that the above problem is non-convex, since it involves the product term of $\bm e$ and $\bm \Phi$. In the following subsection, we will introduce a CGD method to solve the non-convex optimization problem \eqref{eq:SDP1}.

\subsection{Conjugate Gradient Descent Algorithm}

Define ${\bm U_m} = {\mathbb{T}}(\bm Q_m)  \in \mathbb{C}^{N_bN_d \times N_bN_d} $ and
\begin{eqnarray}
\label{Phi-nu}
{\bm \Theta_m} = \left[ {\begin{array}{*{20}{c}}
	{\bm U_m} & \bm{\phi}_m\\
	{\bm{\phi}_m^H}& {\nu_m}
	\end{array}} \right], ~m=1,...,M. 
\end{eqnarray}
Then problem \eqref{eq:SDP1} is rewritten as
\begin{align}
\label{eq:SDP2}
(\bm{\hat \Phi},\bm{\hat e} ) =&~ \arg \mathop {\min }\limits_{\substack{ {\bm \Theta_m} \in \mathbb{C}^{(N_bN_d+1) \times (N_bN_d+1)} \\ \bm e \in \mathbb{C}^{N_bN_d \times 1} } }  \frac{1}{2}\| \bm Y - {\rm{diag}}(\bm{\hat b} + \bm e) \bm\Phi \|_F^2  + \frac{\gamma}{2N_bN_d} \sum\limits_{m=1}^{M} {\rm{Tr}}({\bm U_m}) +\frac{\gamma}{2} \sum\limits_{m=1}^{M} \nu_m + \eta \| \bm e \|_1, \\
 &~{\rm s.t.}~ {\mathbb{T}}({\mathbb{P}}({\bm U_m})) = {\bm U_m},~ {\bm \Theta_m} \succeq 0,~m=1,...,M, \nonumber
\end{align}
where ${\mathbb{P}}(\cdot)$ denotes an inverse operation on the $N_bN_d\times N_bN_d$ input block Toeplitz matrix, and outputs a $(2N_b-1)\times(2N_d-1)$ matrix. In particular, if we partition the block Toeplitz matrix ${\bm U_m}\in\mathbb{C}^{N_bN_d\times N_bN_d}$ into $N_d \times N_d$ blocks, i.e.,
\begin{eqnarray}
{\bm U_m} = \left[ {\begin{array}{*{20}{c}}
	{{\bm{\bar{U}}}_{m,1,1}} & {{\bm{\bar{U}}}_{m,1,2}} & \cdots & {{\bm{\bar{U}}}_{m,1,N_d}}\\
	{{\bm{\bar{U}}}_{m,2,1}} & {{\bm{\bar{U}}}_{m,2,2}} & \cdots & {{\bm{\bar{U}}}_{m,2,N_d}}\\
	\vdots & \vdots & \ddots & \vdots \\
	{{\bm{\bar{U}}}_{m,N,1}} & {{\bm{\bar{U}}}_{m,N,2}} & \cdots & {{\bm{\bar{U}}}_{m,N_d,N_d}}
	\end{array}} \right] \in \mathbb{C}^{N_bN_d \times N_bN_d},
\end{eqnarray}
then the $(i,j)$-th element of ${\mathbb{P}}({\bm U_m})$ is given by
\begin{eqnarray}
\label{eq:PUij}
{\mathbb{P}}({\bm U_m})(i,j) = \frac{1}{\beta_{i,j}} \sum_{d1-d2=i}^{b_1-b_2=j} {{\bm{\bar{U}}}_{m,d_1,d_2}}(b_1,b_2),~b_1,b_2 = 1,2,...,N_b;~d_1,d_2=1,2,...,N_d,
\end{eqnarray}
where
\begin{align}
{\beta_{i,j}} =&~ (N_b-| j |)(N_d-| i |),~j = -N_b + 1, -N_b + 2, ..., N_b-1,~i = -N_d + 1, -N_d + 2, ..., N_d-1. 
\end{align}
%\begin{align}
%{\beta_{i,j}} =&~ (N_b-| j |)(N_d-| i |),\\
%&~j = -N_b + 1, -N_b + 2, ..., N_b-1,\nonumber \\
%&~i = -N_d + 1, -N_d + 2, ..., N_d-1. \nonumber 
%\end{align}

To solve \eqref{eq:SDP2} via the CGD algorithm, we need relax it to a smooth unconstrained form. Hence, we first replace the constraint ${\mathbb{T}}({\mathbb{P}}({\bm U_m})) = {\bm U_m},~m=1,...,M$ by the penalty term $\sum\limits_{m=1}^{M} \frac{\rho}{2}\| {\mathbb{T}}({\mathbb{P}}({\bm U_m})) - {\bm U_m} \|_F^2$, and approximate the $\ell_1$-norm by a twice continuously differentiable function~\cite{sun2017complete}
\begin{align}
\label{eq:e-approx}
\| \bm e \|_1 \approx&~ \phi_{\varpi}(\bm e) = {\varpi} \sum_{n=1}^{N_dN_b} \log\left( \frac{\exp({|e_n|}/{\varpi})+\exp(-{|e_n|}/{\varpi})}{2} \right) = {\varpi} \sum_{n=1}^{N_dN_b}\log \cosh(\frac{|e_n|}{\varpi}), 
\end{align}
where $e_n$ denotes the $n$-th element in $\bm e$ and ${\varpi}$ is a weight parameter, which controls the smoothing level.

\begin{figure}[!tb]
	\centering

	\subfloat{\includegraphics[width=3.0in]{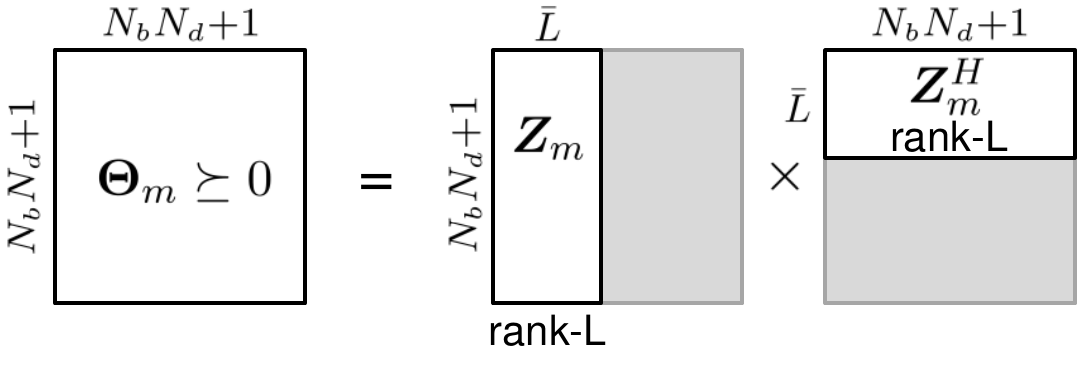}}

	\caption{Principle of non-convex factorization.}
	\label{figure:NF}
\end{figure}

Furthermore,  we remove the constraints ${\bm \Theta_m} \succeq 0,~m=1,...,M$ by setting ${\bm \Theta_m} = \bm{Z}_m \bm{Z}_m^H$ with $\bm{Z}_m \in \mathbb{C}^{(N_bN_d+1) \times \bar L}$, such that $\bar L$ is chosen minimally according to the ranks of ${\bm \Theta_m},~m=1,...,M$. In particular, we have the following lemma. The proof is given in Appendix A.
\begin{lemma}\label{lemma1}
	Suppose $\bm{\hat \Phi}$ is the solution to \eqref{eq:optimization-problem}, where
\begin{eqnarray}
\label{eq:Phi-solution1}
\bm{\hat\phi}_m = \sum_{\ell=1}^{L}\hat c_{\ell,m} \bm a(\hat \tau_{\ell,m},\hat f_{\ell,m}),~m=1,...,M.
\end{eqnarray}
Then each $\bm{\hat\Theta}_m$ in the solution to \eqref{eq:SDP2} is rank-$L$ if $N_bN_d\geq1025$ {\footnote{Actually, the condition $N_bN_d\geq1025$ is a technical requirement that used in Theorem 1 in \cite{chi2015compressive}. Through simulations we found that the result still holds without such condition.}} and $\Delta_m^{\tau,f} = \min_{\ell_1 \neq \ell_2}\max\{|\tau_{\ell_1,m}-\tau_{\ell_2,m}|,|f_{\ell_1,m}-f_{\ell_2,m}|\} \geq \frac{4.76}{N_bN_d},~m=1,...,M$ hold.
\end{lemma}

The above Lemma~\ref{lemma1} shows that each ${\bm \Theta_m}$ in the solution to \eqref{eq:SDP2} should be rank-$L$. It is mentioned in~\cite{burer2003nonlinear} that an algorithm can be accelerated through non-convex factorization if the solution is low-rank, since the size of the optimization variables is significantly reduced (see Fig.~\ref{figure:NF}). Hence, let $\bar L$ be the upper bound on the number of reflectors. We can then let ${\bm \Theta_m} = \bm{Z}_m\bm{Z}_m^H$ such that the constraints ${\bm \Theta_m} \succeq 0,~m=1,...,M$ and ${\rm rank}({\bm \Theta_m}) \leq \bar L$ are both satisfied. Then, \eqref{eq:SDP2} can be rewritten as the following smooth unconstrained optimization problem
\begin{eqnarray}
\label{eq:SDP3}
\min_{\substack{ \bm{Z}_m \in \mathbb{C}^{(N_bN_d+1) \times \bar L } \\ \bm e \in \mathbb{C}^{N_bN_d \times 1} } }   \zeta\left( \{\bm{Z}_m \bm{Z}_m^H\}_{m=1}^M,\bm e\right),
\end{eqnarray}
where
\begin{align}
\label{eq:zeta}
\zeta\left( \{{\bm Z_m} {\bm Z_m^H}\}_{m=1}^M,\bm e\right) =&~  \frac{1}{2}\| \bm Y - {\rm{diag}}(\bm{\hat b} + \bm e) \bm\Phi \|_F^2  + \frac{\gamma}{2N_bN_d} \sum\limits_{m=1}^{M} {\rm{Tr}}({\bm U_m}) \nonumber \\
&~+\frac{\gamma}{2} \sum\limits_{m=1}^{M} \nu_m + \eta \phi_{\varpi}(\bm e) + \sum\limits_{m=1}^{M} \frac{\rho}{2}\| {\mathbb{T}}({\mathbb{P}}({\bm U_m})) - {\bm U_m} \|_F^2.
\end{align}

The CGD algorithm for solving \eqref{eq:SDP3} performs the following iterations
\begin{align}
\label{eq:CG-iterations1}
{\bm Z}_m^i = &~{\bm Z}_m^{i-1} + \mu^i {\bm G}_m^i,~m=1,...,M, \\
\label{eq:CG-iterations2}
\bm e^i = &~\bm e^{i-1} + \mu^i \bm g^i,
\end{align}
where $\mu^i$ is the step size, which is chosen according to the backtracking line search~\cite{bertsekas1999nonlinear}, given in Appendix B, to guarantee that the objective function does not increase with $i$; ${\bm G}_m^i \in \mathbb{C}^{(N_bN_d+1) \times \bar L}$ and $\bm g^i \in \mathbb{C}^{N_bN_d \times 1}$ are the search directions
\begin{align}
\label{gradient-G}
{\bm G}_m^i = &~ -\nabla_{{\bm Z_m}}^i\zeta + \bar\mu^i {\bm G}_m^{i-1}, \\
\label{gradient-g}
\bm g^i = &~ -\nabla_{\bm e}^i\zeta + \bar\mu^i \bm g^{i-1},
\end{align}
with ${\bm G}_m^0 = -\nabla_{{\bm Z_m}}^0\zeta$ and $\bm g^0 = -\nabla_{\bm e}^0\zeta$, where $\nabla_{{\bm Z_m}}^i\zeta$ and $\nabla_{\bm e}^i\zeta$ are the gradients, which are derived in Appendix C; and
\begin{align}
\bar\mu^i = \frac{ \sum\limits_{m=1}^{M} \langle \nabla_{{\bm Z_m}}^i\zeta, \nabla_{{\bm Z_m}}^i\zeta - \nabla_{{\bm Z_m}}^{i-1}\zeta \rangle + \langle \nabla_{\bm e}^i\zeta, \nabla_{\bm e}^i\zeta - \nabla_{\bm e}^{i-1}\zeta \rangle}{ \sum\limits_{m=1}^{M} \langle {\bm G}_m^{i-1}, \nabla_{{\bm Z_m}}^i\zeta - \nabla_{{\bm Z_m}}^{i-1}\zeta \rangle + \langle \bm g^{i-1}, \nabla_{\bm e}^i\zeta - \nabla_{\bm e}^{i-1}\zeta \rangle},
\end{align}
where $\langle \bm X,\bm Y \rangle = {\rm Tr}(\bm Y^H \bm X)$. The iterations in \eqref{eq:CG-iterations1} and \eqref{eq:CG-iterations2} stop when $\sum_{m=1}^{M} \| \nabla_{{\bm Z_m}}^i\zeta \|_F + \|\nabla_{\bm e}^i\zeta\|_2< \epsilon$, where $\epsilon$ is the error tolerance.

\begin{algorithm}[!h] \small
	\label{tab:A1}
	\caption{Conjugate Gradient Descent Algorithm for Solving \eqref{eq:SDP3}.}
	\begin{tabular}{lcl}
		Input $\bm{Y}, \bm{\bar b}$, $N_b$, $N_d$, $\bar L$, $M$, $\gamma$, $\eta$, $\epsilon$, $\rho$ and ${\varpi}$.\\
		1, Initialize $\bm{Z}_m^0$ and $\bm e^0$ as random variables, ${\bm G}_m^0 = -\nabla_{\bm{Z}_m}^0\zeta$, $\bm g^0 = -\nabla_{\bm e}^0\zeta$, and $i=1$. \\
		\sf{Repeat} \\
		\hspace{0.4cm} 2, Calculate $\nabla_{\bm e}^i\zeta$ according to \eqref{nabla-e-1} and \eqref{nabla-e-2}. \\
		\hspace{0.4cm} 3, Calculate $\nabla_{\bm{Z}_m}^i\zeta$ according to \eqref{nabla-Z-1}. \\

		\hspace{0.4cm} 4, Calculate ${\bm G}_m^i$ and $\bm g^i$ according to \eqref{gradient-G} and \eqref{gradient-g}. \\

		\hspace{0.4cm} 5, Obtain $\mu^i$ via backtracking line search in Algorithm 2.\\
		\hspace{0.4cm} 6, Calculate $\bm{Z}_m^i$ and $\bm e^i$ according to \eqref{eq:CG-iterations1} and \eqref{eq:CG-iterations2}.\\
		\hspace{0.4cm} 7, $i = i+1$.\\
		\sf{Until $\| \nabla_{\bm{Z}_m}^i\zeta \|_F + \|\nabla_{\bm e}^i\zeta\|_2 < \epsilon$.} \\
		8, $\bm{\hat e} = \bm e^i$ and obtain $\bm{\hat \Phi}$ according to \eqref{Phi-nu} with ${{\bm{\hat{\Theta}}_m}} = \bm{Z}_m^i (\bm{Z}_m^i)^H$. \\
%		\midrule
%		Return $\bm{\hat\Phi}$ and $\bm{\hat e}$.\\
	\end{tabular}
\end{algorithm}

After \eqref{eq:SDP3} is solved, $\bm\Phi$ can be obtained according to \eqref{Phi-nu}. Note that the unknown number of targets $L$, delays $\bm\tau_m$, Doppler shifts $\bm f_m$ and complex gains $\bm c_m$ in $\bm\Phi$ can then be easily determined by using the two-dimensional MUSIC (2D-MUSIC)~\cite{zheng2017super,berger2010signal} algorithm with each $\bm \phi_m$ as an input. In particular, the 2D-MUSIC method estimates the delays and Doppler shifts of targets by locating the poles in the spectrum and estimates the complex gains by the least-squares method with the estimated delays and Doppler shifts. We refer the readers to standard treatments in~\cite{zheng2017super,berger2010signal} for more details. For clarity, we summarize the proposed CGD method in Algorithm 1. The computational complexity of the proposed algorithm at each iteration is mainly determined by the calculation of $\bm{Z}_m \bm{Z}_m^H$ and the gradient in \eqref{nabla-Z-1}, whose complexity is ${\cal O}(N_b^2N_d^2 \bar L M)$. Note that when the bit-error-rate (BER) in $\bm{\hat b}$ is large, we can perform iterative demodulation at the radar receiver side to improve the performance: in each iteration, after solving \eqref{eq:SDP3}, we make use of $\bm{\hat e}$ and the current $\bm{\hat b}$ to obtain a refined demodulation
\begin{eqnarray}
\label{eq:demodulation}
\bm{\tilde b} = \arg\min_{\bm b\in{\cal B}^{N_d}}\| \bm b - \bm{\hat b} - \bm{\hat{e}}\|_2,
\end{eqnarray}
where ${\cal B}$ is the modulation symbol constellation set. Then we update $\bm{\hat b} \leftarrow \bm{\tilde b}$ in \eqref{eq:SDP3} and solve \eqref{eq:SDP3} again.

\section{Stage 2: Position-velocity Estimation}

In this section, based on the estimated delays and Doppler shifts obtained in Section III, two target position-velocity estimation methods are discussed. Both utilize the delay differences between different receivers to calculate the target positions. One is based on solving a set of nonlinear equations, and the other is based on BP neural network.

\subsection{Estimator Based on Solving Nonlinear Equations}

From \eqref{eq:tau} and \eqref{eq:tau-DeltaM}, we have
\begin{eqnarray}
\label{eq:tau-2}
{\tau}_{\ell,m} = \frac{{\bar\tau}_{\ell,m} - \Delta_{\tau}}{N_dT} = \frac{1}{N_dT} \left( \frac{1}{c}(\|\bm p_0 - \bm x_{\ell}\|_2 + \| \bm p_m - \bm x_{\ell}\|_2) - \Delta_{\tau} \right).
\end{eqnarray}
For tracking multi-moving targets, their Doppler shifts and complex gains are different, which help us to distinguish the delays of targets at different receivers. When these delays can be distinguished, we use the delay difference at different receivers to eliminate $\|\bm p_0 - \bm x_{\ell}\|_2$ and $\Delta_{\tau}$ to determine $\bm x_{\ell}$, i.e.,
\begin{eqnarray}
\label{eq:tau-3}
cN_dT({\tau}_{\ell,m} - {\tau}_{\ell,1}) =  \| \bm p_{m} - \bm x_{\ell}\|_2 -  \| \bm p_1 - \bm x_{\ell}\|_2,~m = 2,...,M.
\end{eqnarray}
Denote
\begin{eqnarray}
\label{eq:Ytau-set}
{\cal Y}({\tau}_{\ell,m},{\tau}_{\ell,1}, \bm p_m,  \bm p_1, \bm x_{\ell}) = cN_dT({\tau}_{\ell,m} - {\tau}_{\ell,1}) -  \| \bm p_{m} - \bm x_{\ell}\|_2 +  \| \bm p_1 - \bm x_{\ell}\|_2,~m = 2,...,M.
\end{eqnarray}
Then we get $M-1$ equations and the target position can be estimated by solving three equations at a time{\footnote{Note that due to the estimation errors of ${\tau}_{\ell,m}$, directly solving $M-1$ equations in \eqref{eq:Ytau-set} usually leads to infeasibility. Hence we divide them into multiple groups and solve only three equations at a time.}}, i.e., by solving
\begin{align}
\label{eq:equationset}
\begin{cases}
{\cal Y}({\hat\tau}_{\ell,m},{\hat\tau}_{\ell,1}, \bm p_m,  \bm p_1, \bm x_{\ell}) = 0, \\
{\cal Y}({\hat\tau}_{\ell,m+1},{\hat\tau}_{\ell,1}, \bm p_{m+1},  \bm p_1, \bm x_{\ell}) = 0, \\
{\cal Y}({\hat\tau}_{\ell,m+2},{\hat\tau}_{\ell,1}, \bm p_{m+2},  \bm p_1, \bm x_{\ell}) = 0, \\
\bm x_{\ell} \in {\cal S},
\end{cases}
~m=2,...,M-2, 
\end{align}
where ${\cal S}$ denotes the surveillance area. The above equation sets can be solved with some numerical solvers, e.g., the ``solver'' function in Matlab, and we named this method as the ``solver'' method. The final estimated target position $\bm{\hat x}_{\ell} = [\hat x^x_{\ell},\hat x^y_{\ell},\hat x^z_{\ell}]$ can be obtained by averaging the $M-3$ solutions to \eqref{eq:equationset}.

Once the estimate $\bm{\hat x}_{\ell}$ is available by solving \eqref{eq:equationset} and the estimate $\hat f_{\ell,m}$ is obtained from Section III, from \eqref{eq:doppler} and \eqref{eq:tau-DeltaM} we have
\begin{align}
\label{eq:doppler2}
f_{\ell,m} = {\bar f}_{\ell,m}{NT} = \frac{NT}{\lambda} \left( \frac{\bm p_0 - \bm x_{\ell}}{ \| \bm p_0 - \bm x_{\ell}\|_2} + \frac{\bm p_m - \bm x_{\ell}}{ \| \bm p_m - \bm x_{\ell}\|_2} \right)^T \bm v_{\ell},~m=1,...,M.
\end{align}
Then the velocity can be easily determined by
\begin{align}
\label{eq:velocity-est}
\bm{\hat v}_{\ell} = \bm\Gamma_{\ell}^{\dagger} \bm{\hat f}_{\ell},
\end{align}
where $(\cdot)^{\dagger}$ denotes the pseudo-inverse, i.e., $\bm Y^{\dagger} = (\bm Y^H \bm Y)^{-1}\bm Y^H$, the $m$-th element of $\bm{\hat f}_{\ell} \in \mathbb{R}^{M\times 1}$ is $\hat f_{\ell,m}$ and the $m$-th row of $\bm\Gamma_{\ell} \in \mathbb{R}^{M\times 3}$ is
\begin{align}
\frac{NT}{\lambda} \left( \frac{\bm p_0 - \bm{\hat x}_{\ell}}{ \| \bm p_0 - \bm{\hat x}_{\ell}\|_2} + \frac{\bm p_m - \bm{\hat x}_{\ell}}{ \| \bm p_m - \bm{\hat x}_{\ell}\|_2} \right)^T.
\end{align}

\subsection{Estimator Based on Neural Network}

\begin{figure}%[!htb]
	\centering

	\subfloat{\includegraphics[width=2.1in]{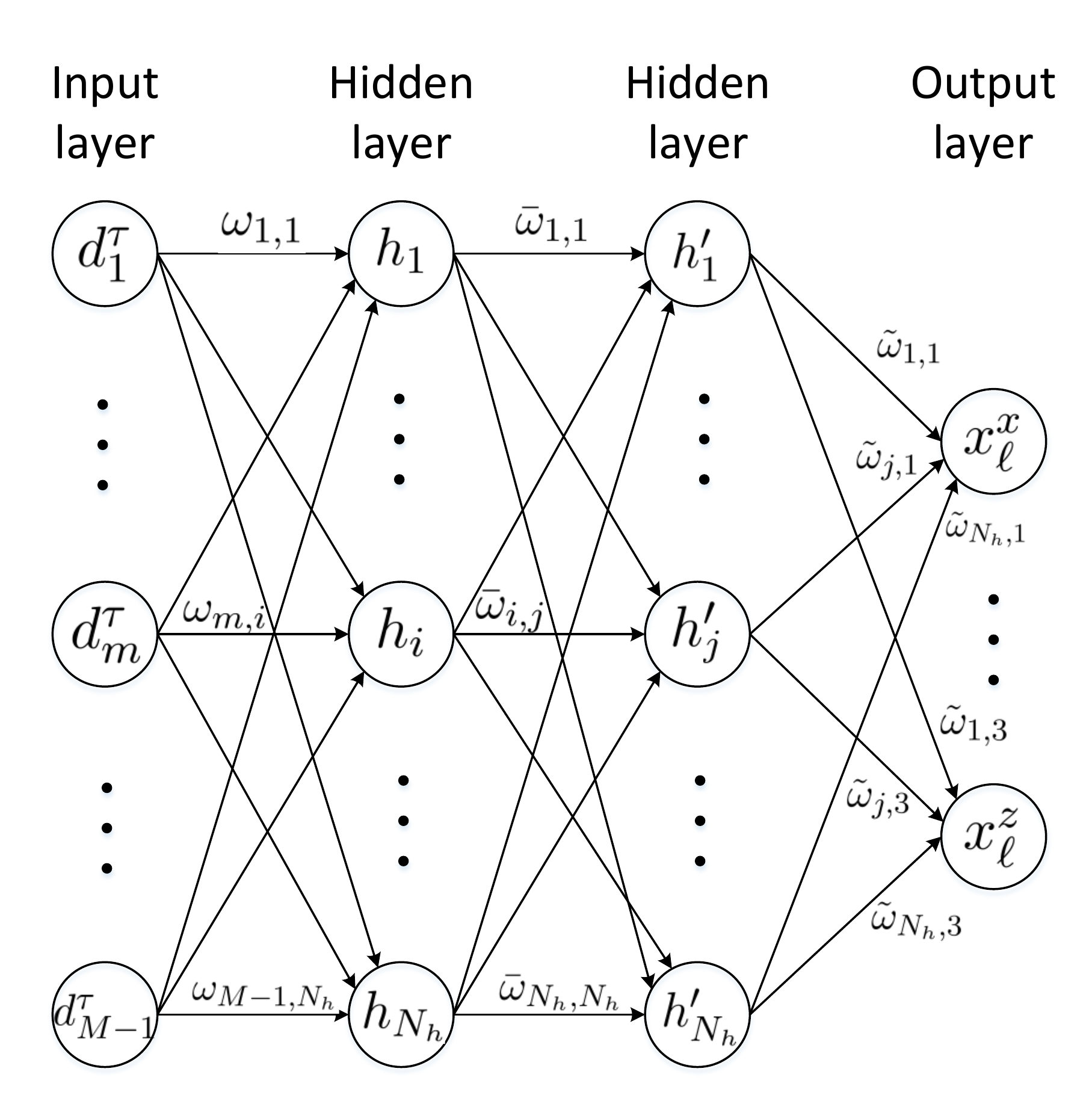}}

	\caption{BP neural network for position estimation.}
	\label{figure:BP}
\end{figure}

In~\cite{yang2018fast}, a sound source angle estimation method based on neural network is proposed, which has better performance and lower computational complexity compared with the traditional method of solving nonlinear equations. This work inspired us to use a BP neural network to estimate the target position.

As shown in Fig.~\ref{figure:BP}, the proposed BP neural network is composed of four layers, namely, an input layer, two hidden layers{\footnote{Note that we use two hidden layers here, although according to Kolmogorov's theorem~\cite{kuurkova1992kolmogorov}, a three-layer BP neural network can approximate an arbitrary nonlinear function with arbitrary accuracy. Through simulation we found that by using two hidden layers, we can use fewer neurons than using one hidden layer to achieve the same performance, and the training time is also less.}}, and an output layer. The input layer has $M-1$ neurons, corresponding to $M-1$ delay differences $\bm d_{\ell}^{\tau} = cN_dT[({\tau}_{\ell,2} - {\tau}_{\ell,1}),...,({\tau}_{\ell,M} - {\tau}_{\ell,1})] \in \mathbb{R}^{(M-1)\times 1}$. Each hidden layer has $N_h$ neurons, and the output layer has 3 neurons, corresponding to the target coordinates $\bm x_{\ell} = [x_{\ell}^x, x_{\ell}^y, x_{\ell}^z]$. Define the output of two hidden layers as 
$\bm h = [h_1,...,h_{N_h}] \in \mathbb{R}^{N_h\times 1}$ and $\bm h' = [h'_1,...,h'_{N_h}] \in \mathbb{R}^{N_h\times 1}$, respectively. Then the mapping functions between two adjacent layers are given by
\begin{align}
h_i =&~ {\cal L}_1(\sum_{m=1}^{M-1}\omega_{m,i}d^{\tau}_m + \psi_i), \\
h'_j =&~ {\cal L}_2(\sum_{i=1}^{N_h}\bar\omega_{i,j}h_i + \bar{\psi}_j), \\
x_n = &~ {\cal L}_3(\sum_{j=1}^{N_h}\tilde\omega_{j,n}h'_j + \tilde{\psi}_n),
\end{align}
where $d^{\tau}_m$, $h_i$, $h'_j$ and $x_n$ are the $m$-th, $i$-th, $j$-th and $n$-th elements of $\bm d^{\tau}$, $\bm h$, $\bm h'$ and $\bm x_{\ell}$, respectively; $\omega_{m,i}$, $\bar\omega_{i,j}$, $\tilde\omega_{j,n}$ are the connection weights between neurons (see Fig.~\ref{figure:BP}); $\psi_i$, $\bar{\psi}_j$, $\tilde{\psi}_n$ are the activation thresholds of the corresponding neuron; ${\cal L}_1(\cdot)$, ${\cal L}_2(\cdot)$ and ${\cal L}_3(\cdot)$ represent the activation functions of the first and second hidden layer and output layer neurons, respectively, which are set as the ``tanh'' function:
\begin{align}
{\cal L}(x) = \frac{e^x-e^{-x}}{e^x+e^{-x}}.
\end{align}

During the training phase, the reflector positions are randomly generated, and then according to \eqref{eq:tau-3}, the corresponding delay differences are calculated. These simulated delay differences and reflector positions are respectively used as the input and output for neural network training. After the neural network is well trained, for each estimated delay difference vector $\bm d_{\ell}^{\tau}$ as input, it outputs the location estimate $\bm{\hat x}_{\ell}$. The velocity can then be determined by \eqref{eq:velocity-est}.

\section{Simulation Results}

\subsection{Baseline for Comparison: Convex Relaxation Method}

As a baseline of comparison, we consider a convex relaxation (CR) method for estimating the continuous delays and Doppler shifts in Section III. That is, since estimating the unknown $\bm e$ and $\bm \Phi$ from their product term is non-convex, we ignore the error $\bm e$ in \eqref{eq:problem} as
\begin{eqnarray}
\label{eq:problem-reduce}
\bm Y = {\rm{diag}}(\bm{\hat b}) \bm\Phi  + \bm W.
\end{eqnarray}
Then $\bm\Phi$ can be determined by solving the following optimization problem
\begin{align}
%\label{eq:optimization-problem-re}
\bm{\hat\Phi} =&~ \arg \mathop {\min }\limits_{\substack{ \bm\Phi \in \mathbb{C}^{N_bN_d \times M} } }  \frac{1}{2}\| \bm Y - {\rm{diag}}(\bm{\hat b}) \bm\Phi \|_F^2  + \bar\gamma \sum_{m=1}^{M}\| \bm\phi_m \|_{{\cal A}}, \nonumber \\
\label{eq:SDP-re}
=&~ \arg \mathop {\min }\limits_{\substack{ \bm\Phi \in \mathbb{C}^{N_bN_d \times M},\\ \bm Q_m \in \mathbb{C}^{(2N_b-1) \times (2N_d-1)} } }  \frac{1}{2}\| \bm Y - {\rm{diag}}(\bm{\hat b} ) \bm\Phi \|_F^2  + \frac{\bar\gamma}{2N_bN_d} \sum\limits_{m=1}^{M} {\rm{Tr}}({\mathbb{T}}(\bm Q_m)) +\frac{\bar\gamma}{2}\sum\limits_{m=1}^{M} \nu_m, \\
 &~{\rm s.t.}~\left[ {\begin{array}{*{20}{c}}
	{\mathbb{T}}(\bm Q_m)& \bm{\phi}_m\\
	{\bm{\phi}_m^H}& {\nu_m}
	\end{array}} \right] \succeq 0,~m=1,...,M, \nonumber
\end{align}
where $\bar\gamma$ is a weight factor. Problem \eqref{eq:SDP-re} does not take into account the BER and is convex, hence it can be solved with standard convex solvers, e.g., CVX~\cite{boyd2004convex}. And the complexity in each iteration is ${\cal O}(N_b^6N_d^6 M)$ if the interior point method is used~\cite{zheng2018adaptive}.

\subsection{Simulation Setup}

\subsubsection{Basic parameter setting}
In order to demonstrate the performance of the proposed algorithms, we simulate a scenario of having several paths reflected by reflectors between an OFDM transmitter and $M=4$ radar receivers. The carrier frequency is 2 GHz. Let $N_b = 16$, $N_d = 16$, $N_p = 16$ and the total bandwidth be $320$ kHz, i.e., the frequency spacing between adjacent subcarriers is $20$ kHz. Hence the duration of data symbols is $N_dT = 50 \mu\text{s}$ and the duration of CP is $50 \mu\text{s}$, so the block length is $100 \mu\text{s}$ and the time of collecting $16$ data blocks is $1.6 \text{ms}$.

The transmitted OFDM signal is generated according to \eqref{eq:st} with normalized quadrature phase-shift keying (QPSK) data symbols. Both the targets and clutters are assumed to be point scatterers in our simulations. For simplicity, the complex path gains $\{c_{\ell,m}\}$ are generated with fixed magnitude $c_0$ and random phases. Based on \eqref{eq:ynm2}, we define the SNR at the radar receivers as
\begin{align}
\text{SNR}= \frac{ \mathbb{E}\{|  \sum_{\ell = 1}^{L} c_{\ell,m} e^{i2\pi n f_{\ell,m}} {e^{ - i2\pi k \tau_{\ell,m}}} |^2 \} }{\sigma_w^2} = \frac{ \sum_{\ell = 1}^{L} \mathbb{E}\{|  c_{\ell,m} |^2 \} }{\sigma_w^2} = \frac{ L c_0^2 }{\sigma_w^2},
\end{align}
%\begin{align}
%\text{SNR}=&~ \frac{ \mathbb{E}\{|  \sum_{\ell = 1}^{L} c_{\ell,m} e^{i2\pi n f_{\ell,m}} {e^{ - i2\pi k \tau_{\ell,m}}} |^2 \} }{\sigma_w^2} \nonumber \\
%=&~ \frac{ \sum_{\ell = 1}^{L} \mathbb{E}\{|  c_{\ell,m} |^2 \} }{\sigma_w^2} \nonumber \\ 
%=&~ \frac{ L c_0^2 }{\sigma_w^2},
%\end{align}
where $\sigma_w^2$ is the variance of the Gaussian noise sample $\bar w_{n,m}(k)$ in \eqref{eq:ynm2}.
In addition, when the demodulation error is considered, the mistaken demodulation is controlled by the BER.
 
The transmitter position and receiver positions are respectively set as $\bm p_0 = [5\text{km},300\text{m},200\text{m}]^T$, $\bm p_1 = [0\text{m},0\text{m},0\text{m}]^T$, $\bm p_2 = [1\text{km},0\text{m},350\text{m}]^T$, $\bm p_3 = [2.5\text{km},0\text{m},1.5\text{km}]^T$ and $\bm p_4 = [4\text{km},0\text{m},780\text{m}]^T$. The surveillance area is in the range of $x^{x}_{\ell}\in(0\text{m},5\text{km}]$, $x^{y}_{\ell}\in(1\text{km},6\text{km}]$, $x^{z}_{\ell}\in(0\text{m},1.5\text{km}]$. The synchronization error is set as $\Delta_{\tau} = 0.1\mu\text{s}$. In this way the maximum path delay satisfies
\begin{align}
N_pT = 50\mu\text{s}>\max_{m,\ell}\{\bar\tau_{\ell,m}-\Delta_{\tau}\} = 45.89 \mu\text{s}.
\end{align}
For neural network training, 4000 training data points are generated according to \eqref{eq:tau-2}. The target positions are uniformly generated on a $20\times 20\times 10$ grid in the surveillance area, i.e., the sampling intervals are $250\text{m}$, $250\text{m}$ and $150\text{m}$ for $x^{x}_{\ell}\in(0\text{m},5\text{km}]$, $x^{y}_{\ell}\in(1\text{km},6\text{km}]$, $x^{z}_{\ell}\in(0\text{m},1.5\text{km}]$, respectively. Each of the two hidden layers has $N_h = 25$ neurons. The maximum epoch of the neural network is 1000 and the training goal is the validation mean squared error equals to $10^{-8}$.%relative $\text{RMSE}_{x} = 10^{-8}$.

For the proposed CGD method, the error tolerance for iterations in \eqref{eq:CG-iterations1} and \eqref{eq:CG-iterations2} is set as $\epsilon = 10^{-6}$. The upper bound in \eqref{eq:SDP3} is set as $\bar L=10$. The weight in \eqref{eq:e-approx} is set as $\varpi = 0.01$. And the weight factors in \eqref{eq:zeta} are set as $\gamma = \sigma_w\sqrt{2\log(N_bN_d)}$, $\eta = \sigma_w\sqrt{N_d\log(N_bN_d)}$ and $\rho = 5$. The weight factor of the CR method in \eqref{eq:SDP-re} is set as $\bar\gamma = \sigma_w\sqrt{2\log(N_bN_d)}$. For the runtime comparisons, the simulations were carried out on an Intel Xeon desktop computer with a 3.5 GHz CPU and 24 GB of RAM.

\subsubsection{Constant-velocity target simulation setting}
In order to quantitatively evaluate the proposed methods, we first consider a constant-velocity target scenario, where targets are randomly generated in the surveillance area and the velocities of targets along $x$, $y$ and $z$-axes are randomly generated between -300 m/s and 300 m/s and fixed.
%We evaluate the relative root-mean-squared-errors (RMSEs) of the target position and velocity estimation for the proposed algorithms. The relative position and velocity RMSEs are respectively calculated as according to the maximum surveillance distance and velocity range
We evaluate the root-mean-squared-error (RMSE) of the estimated $\bm \Phi$ to demonstrate the convergence behavior of the proposed CGD method, which is calculated as $\text{RMSE}_{\Phi}^i = \|\bm \Phi - \bm{\hat\Phi}^{i} \|_F$, where $\bm{\hat\Phi}^{i}$ denotes the estimated $\bm{\Phi}$ in the $i$-th iteration. 
%\begin{align}
%\text{RMSE}_{\Phi}^i =&~ \|\bm \Phi - \bm{\hat\Phi}^{i} \|_F,
%\end{align}

%For the proposed CGD method, when the estimated number of delays is large than a threshold $\tilde L = 10$, we consider the CGD method not convergent enough 
Note that sometimes the 2D-MUSIC algorithm returns a bunch of delays and Doppler shifts (especially for the CR method or BER is large), which can be either true detections or false alarms. For the proposed CGD method, since we consider a case where the number of scatterers is not very large, when the estimated number of delays is large than a threshold $\tilde L = 6$, we consider this is due to the incomplete estimation of the demodulation error and hence perform iterative demodulation by \eqref{eq:demodulation} and solve \eqref{eq:SDP3} again by Algorithm 1{\footnote{Note that we do not always use iterative demodulation because the CGD method can usually converge. Only when the BER is large, the CGD method sometimes may not be able to fully estimate the error in one iteration, and then we use iterative demodulation.}}. In addition, in order to facilitate the evaluation, we assume that $L$ is known and only select the $L$ estimated delays and Doppler shifts with the $L$ largest complex gains $\bm c_m$ for subsequent target position and velocity estimation. Then, we evaluate the relative mean position errors (MPEs) and relative mean velocity errors (MVEs) of target along the $x$, $y$ and $z$-axes respectively as
\begin{align}
\text{RMPE}_{x/y/z} =&~ \frac{1}{\text{MC}} \sum_{{n_{\text{MC}}}=1}^{\text{MC}}  \frac{1}{L}\sum_{\ell=1}^{L} \frac{\left |({x_{\ell}^{x/y/z})}^{n_{\text{MC}}} - {({\hat x_{\ell}^{x/y/z}})}^{n_{\text{MC}}} \right |}{\Delta x^{x/y/z}_{\text{max}}} , \\
\text{RMVE}_{x/y/z} =&~ \frac{1}{\text{MC}} \sum_{{n_{\text{MC}}}=1}^{\text{MC}}  \frac{1}{L}\sum_{\ell=1}^{L} \frac{\left |({v_{\ell}^{x/y/z})}^{n_{\text{MC}}} - {({\hat v_{\ell}^{x/y/z}})}^{n_{\text{MC}}} \right |}{\Delta v^{x/y/z}_{\text{max}}},
\end{align}
where ${\text{MC}}$ is the number of Monte Carlo runs; $({x_{\ell}^{x/y/z})}^{n_{\text{MC}}}$ and $({v_{\ell}^{x/y/z})}^{n_{\text{MC}}}$ are the true position and velocity of the $\ell$-th target along $x/y/z$-axis in the $n_{\text{MC}}$-th run, respectively; while ${({\hat x_{\ell}^{x/y/z}})}^{n_{\text{MC}}}$ and ${({\hat v_{\ell}^{x/y/z}})}^{n_{\text{MC}}}$ are the corresponding estimates, respectively; ${\Delta x^{x}_{\text{max}}} = 5\text{km}$, ${\Delta x^{y}_{\text{max}}} = 5\text{km}$ and ${\Delta x^{z}_{\text{max}}} = 1.5\text{km}$ are the maximum surveillance distance ranges along $x$, $y$ and $z$-axes, respectively; and ${\Delta v^{x}_{\text{max}}} = {\Delta v^{y}_{\text{max}}} = {\Delta v^{z}_{\text{max}}} = 600\text{m/s}$ are the maximum surveillance velocity ranges along $x$, $y$ and $z$-axes, respectively. Note that when the solution of \eqref{eq:equationset} is out of the surveillance area (especially for the CR method), the ``solver'' method returns infeasibility due to the violation of the constraint $\bm x_{\ell} \in {\cal S}$, we hence set
\begin{align}
\frac{\left |{x_{\ell}^{x/y/z}} - {{\hat x_{\ell}^{x/y/z}}} \right |}{\Delta x^{x/y/z}_{\text{max}}} = \frac{\left |{v_{\ell}^{x/y/z}} - {{\hat v_{\ell}^{x/y/z}}} \right |}{\Delta v^{x/y/z}_{\text{max}}} = 1.
%\frac{\left |({x_{\ell}^{x/y/z})}^{n_{\text{MC}}} - {({\hat x_{\ell}^{x/y/z}})}^{n_{\text{MC}}} \right |}{\Delta x^{x/y/z}_{\text{max}}} = \frac{\left |({v_{\ell}^{x/y/z})}^{n_{\text{MC}}} - {({\hat v_{\ell}^{x/y/z}})}^{n_{\text{MC}}} \right |}{\Delta v^{x/y/z}_{\text{max}}} = 1.
\end{align}

%even if when solving three equations at a time in \eqref{eq:equationset}; we set $ \bm{\hat x}_{\ell} = [0\text{m},0\text{m},0\text{m}]$ and $\bm{\hat v}_{\ell} = [0\text{m/s},0\text{m/s},0\text{m/s}]$ in such case. 

%For the If a single/multiple target(s) return(s) a bunch of delays and Doppler shifts, we take delays and Doppler shifts according to the largest one/several complex gains for subsequent target positions and velocities estimation.
%Furthermore, we evaluate the RMSE of the estimated delays as
%\begin{align}
%\text{RMSE}_{\tau} =&~ \sqrt{\frac{1}{\text{MC}} \sum_{{n_{\text{MC}}}=1}^{\text{MC}}  \frac{1}{M}\sum_{m=1}^{M} \frac{1}{L}\sum_{\ell=1}^{L} (\tau_{\ell,m}^{n_{\text{MC}}}-\hat\tau_{\ell,m}^{n_{\text{MC}}})^2 }.
%\end{align}

\subsubsection{Maneuvering target simulation setting}

%Afterwords, we randomly generate a continuously changing velocity of 30s, shown in Fig.~\ref{figure:Scenario-1}(b), and then calculate the target moving path according to its velocity. The blue curve in Fig.~\ref{figure:Scenario-1}(a) shows the target moving path, Then the positions and velocities of targets are generated in the same way as scenario 1, which are shown in Fig.~\ref{figure:Scenario-2}(a) and Fig.~\ref{figure:Scenario-2}(b), respectively. 

%Since the goal is to obtain the target positions and velocities, which is very suitable for tracking the target, 

We simulate two scenarios of tracking maneuvering targets to validate the proposed methods. In the first scenario there are a moving target and two clutters/stationary targets. The initial target position is set as $[500\text{m},4\text{km},1.35\text{km}]$. The target trajectory and velocity are shown in Figs.~\ref{figure:tracking-results-1}(a)-(b) and Fig.~\ref{figure:tracking-S1-V}, respectively. And the positions of clutters are respectively set as $[4.1\text{km},3\text{km},510\text{m}]$ and $[1\text{km},1.5\text{km},300\text{m}]$.  The SNR and BER in scenario 1 are respectively set as 15 dB and 0.01. In the second scenario there are two moving targets. The initial positions of two targets are set as $[500\text{m},1.5\text{km},350\text{m}]$ and $[1.5\text{km},4.5\text{km},1.2\text{km}]$, respectively. And their trajectories and velocities are shown in Fig.~\ref{figure:tracking-results-1}(c) and Fig.~\ref{figure:tracking-S2-V}, respectively. The SNR and BER in scenario 2 are respectively set as 15 dB and 0.03. We assume that the targets need to be tracked for $30\text{s}$ and estimation is performed once per second, i.e., for each second, there is a time interval of $998.4 \text{ms}$ after the initial $1.6 \text{ms}$ of data collection. For tracking multi-moving targets, their Doppler shifts and complex gains are different, which help us to distinguish the delays of targets at different receivers. And we can identify the moving targets and clutters according to their velocities, i.e., a target is considered as a clutter if its velocity $\|\bm v_{\ell}\|_2<3$.

%The blue curves in Fig. \ref{figure:tracking-V1-CGD} and Fig. \ref{figure:tracking-V1-CR} show the ground truths of the target velocities $v_{\ell}^x$, $v_{\ell}^y$ and $v_{\ell}^z$.

\subsection{Performance}

%\subsubsection{Constant-velocity target performance}

In the first simulation, we compare the delay and Doppler shift estimation performances of the proposed CGD and CR methods. The SNR in this simulation is set as 15 dB. The positions and velocities of three targets are respectively set as $\bm x_{1} = [1.8\text{km},5.5\text{km},450\text{m}]$, $\bm v_{1} = [0\text{m/s},0\text{m/s},0\text{m/s}]$, $\bm x_{2} = [800\text{m},1.2\text{km},650\text{m}]$, $\bm v_{2} = [20\text{m/s},80\text{m/s},50\text{m/s}]$ and $\bm x_{3} = [2.5\text{km},3.2\text{km},120\text{m}]$, $\bm v_{3} = [-10\text{m/s},-90\text{m/s},-20\text{m/s}]$. Fig.~\ref{figure:delay-Doppler-results} shows the delay-Doppler estimation results for 4 receivers. We can see that when $\text{BER}=0.01$, the CR method can basically estimate the target delay and Doppler shift with some performance degradation. When considering a higher BER condition ($\text{BER}=0.03$), the CR method returns a large number of false alarms, making it difficult to identify true targets, indicating that the target positions and velocities cannot be determined. In contrast, the proposed CGD method still works well when $\text{BER}=0.03$, and the target delays and Doppler shifts can be clearly determined.

\begin{figure*}[!tb]
	\centering
		
	\subfloat[][]{\includegraphics[width=1.6in]{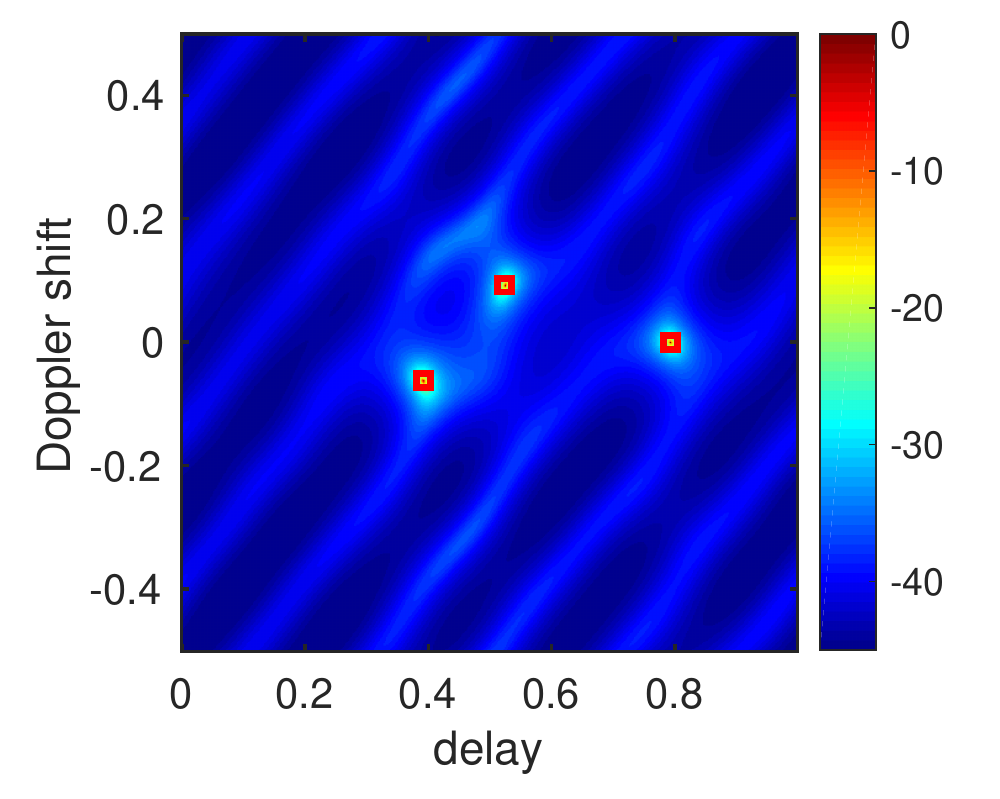}}
	\subfloat[][]{\includegraphics[width=1.6in]{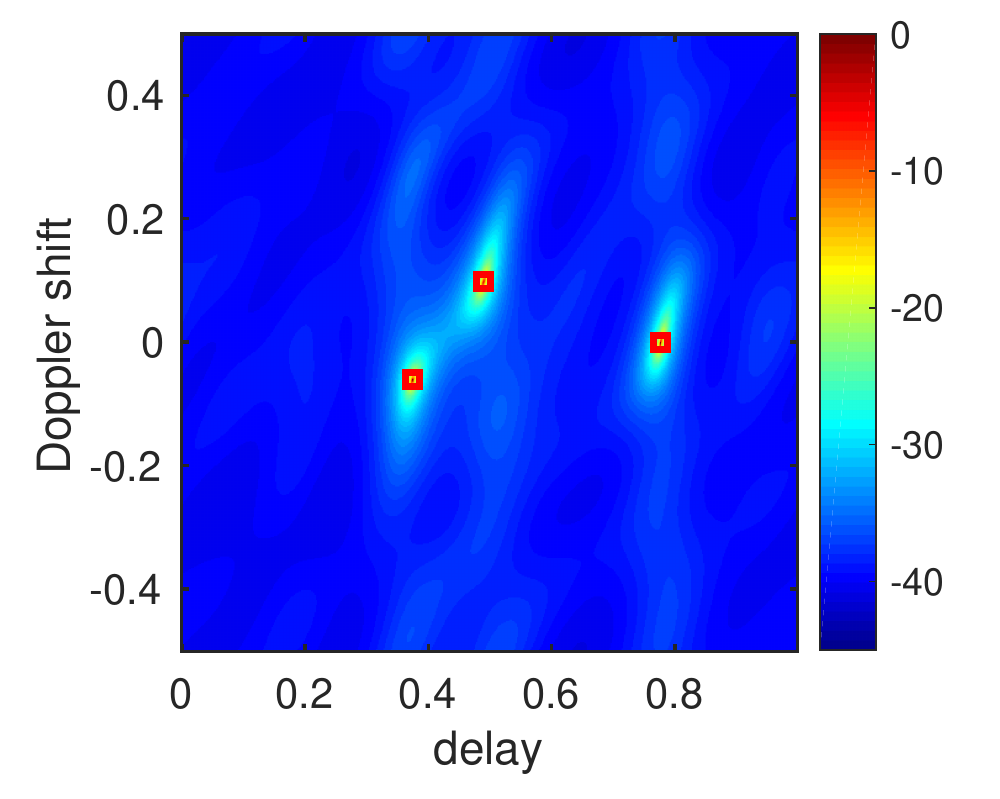}}
	\subfloat[][]{\includegraphics[width=1.6in]{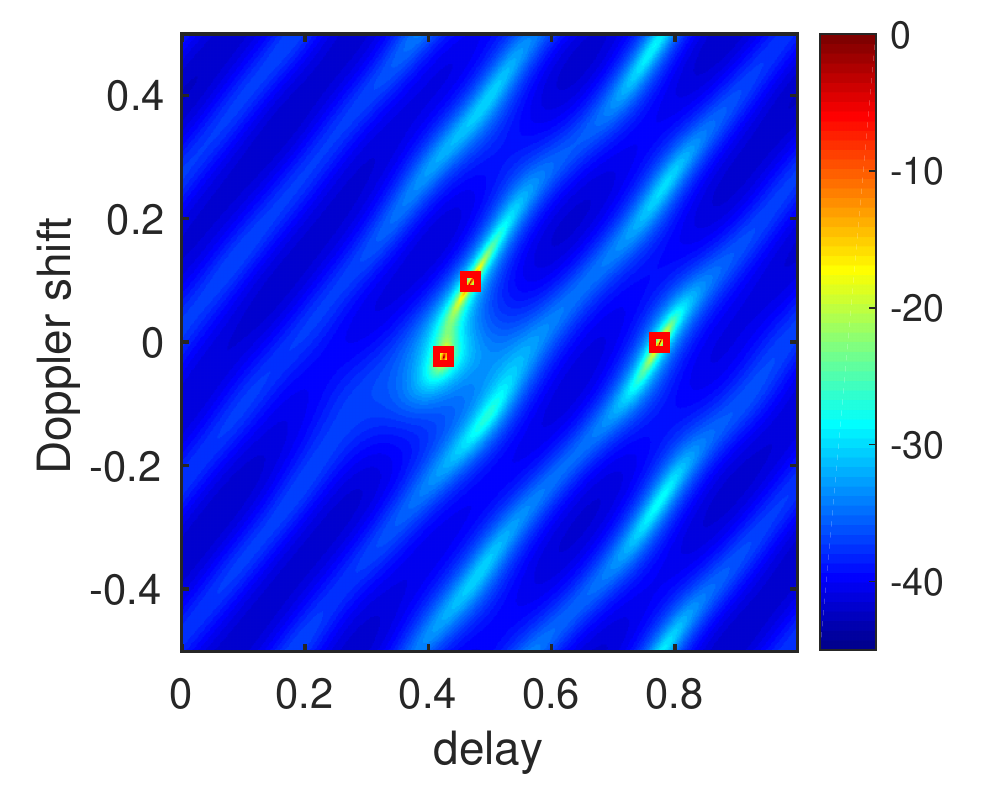}}
	\subfloat[][]{\includegraphics[width=1.6in]{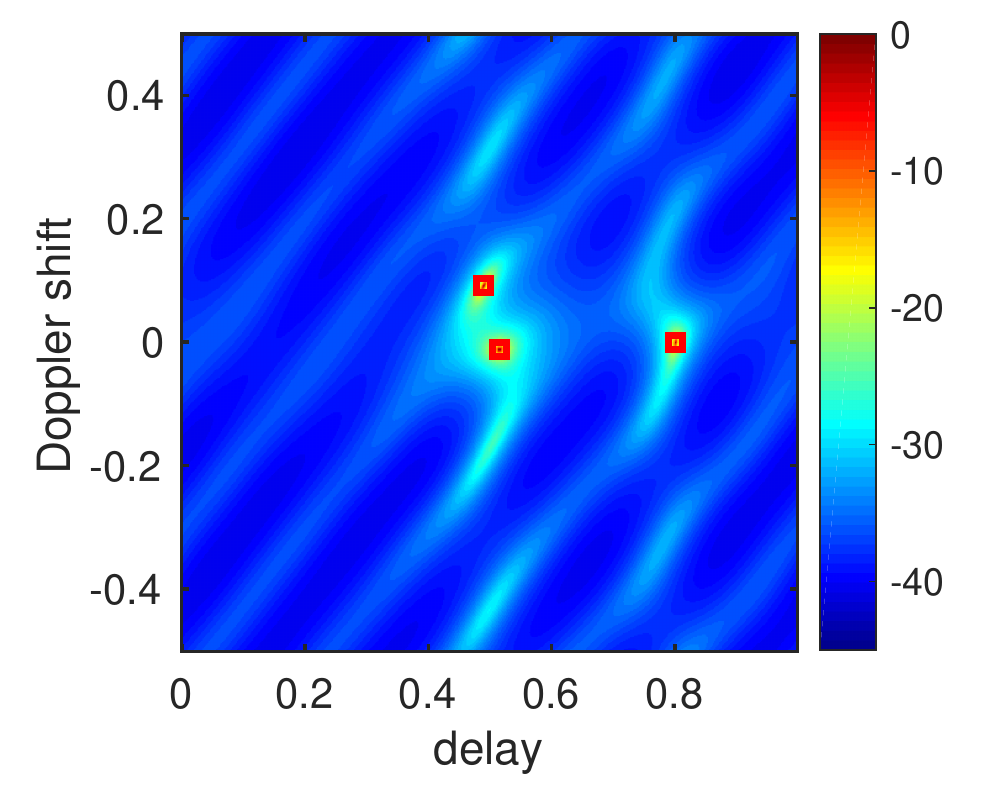}}
	
	\subfloat[][]{\includegraphics[width=1.6in]{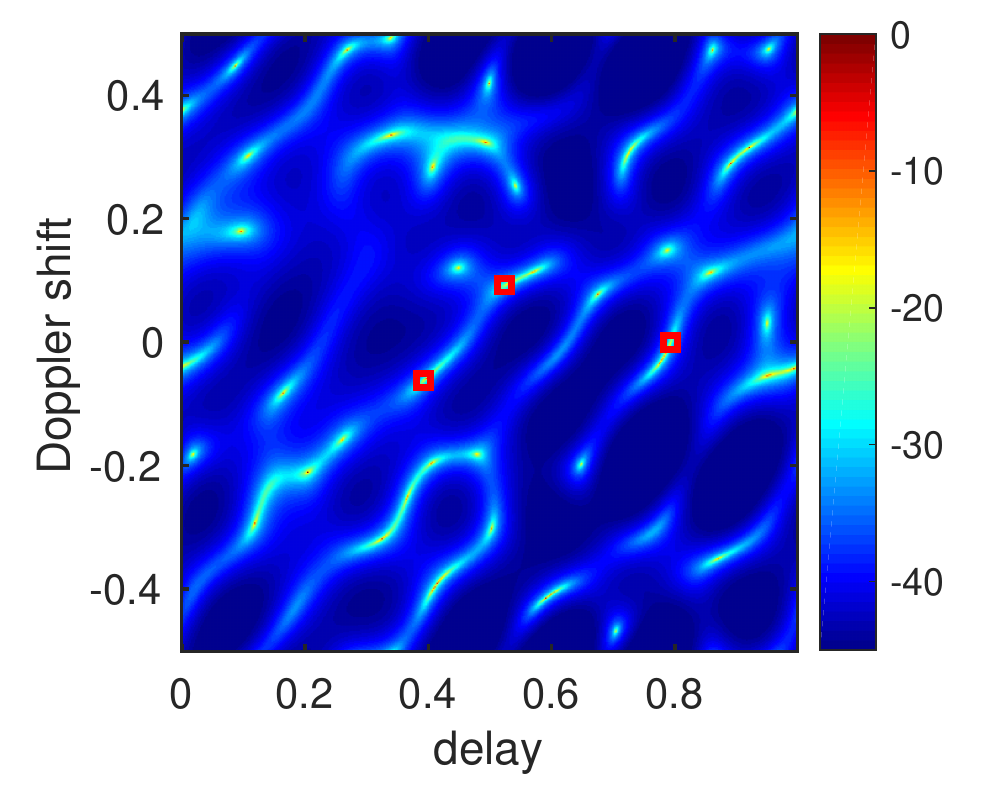}}
	\subfloat[][]{\includegraphics[width=1.6in]{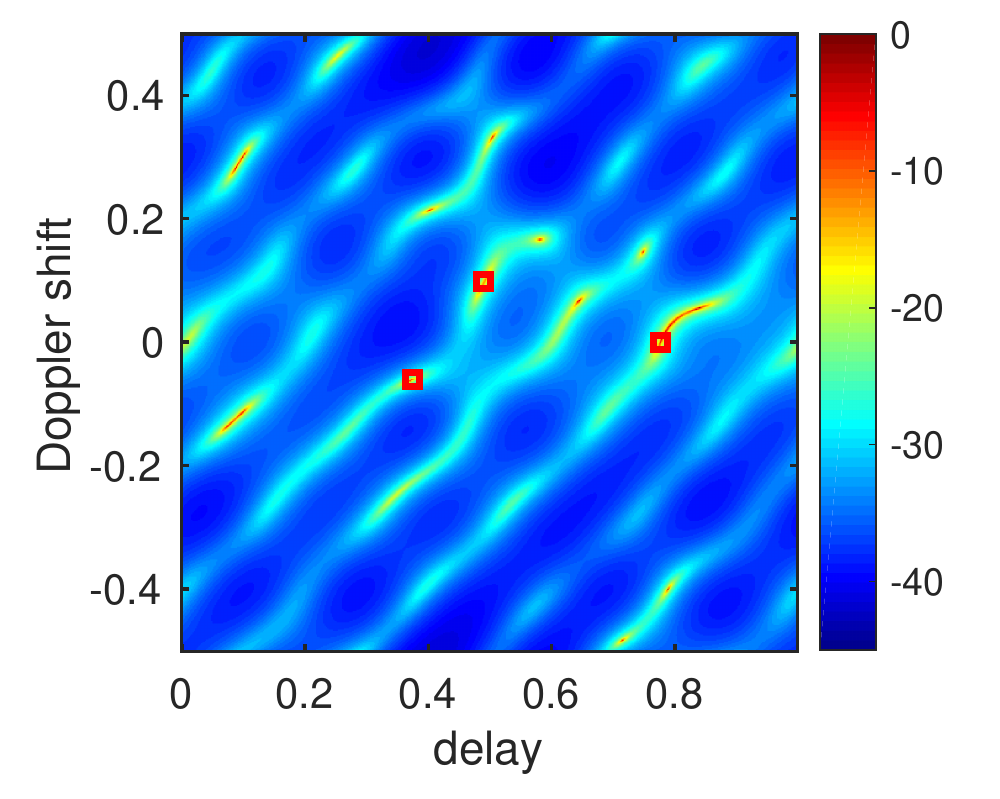}}
	\subfloat[][]{\includegraphics[width=1.6in]{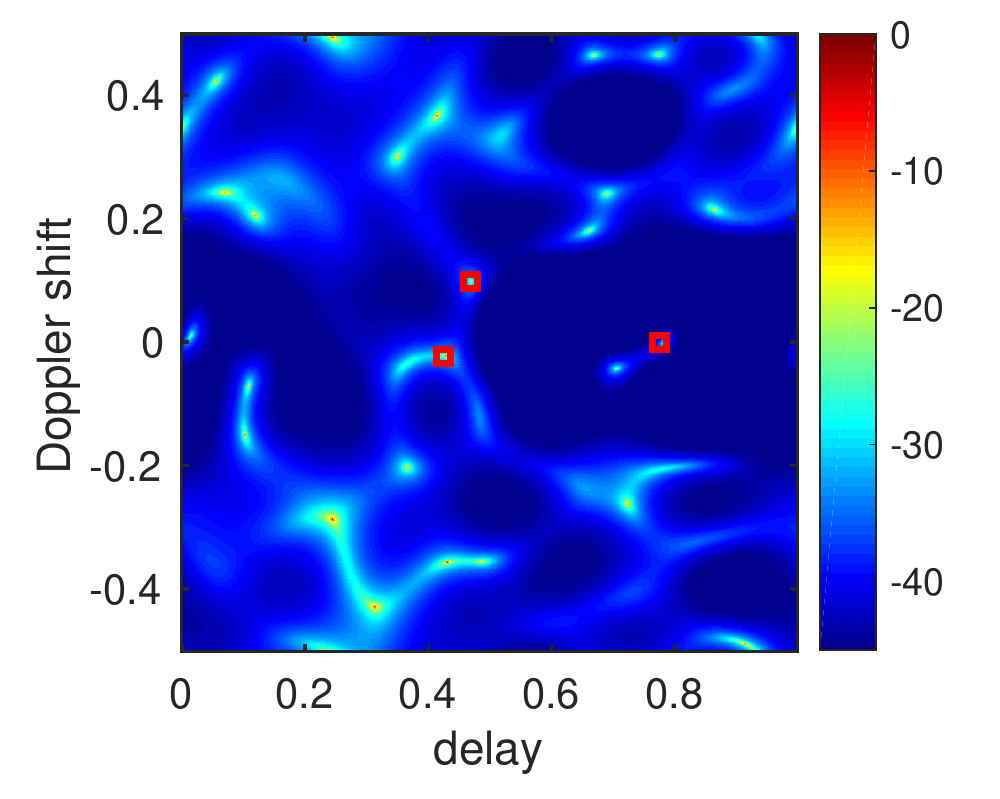}}
	\subfloat[][]{\includegraphics[width=1.6in]{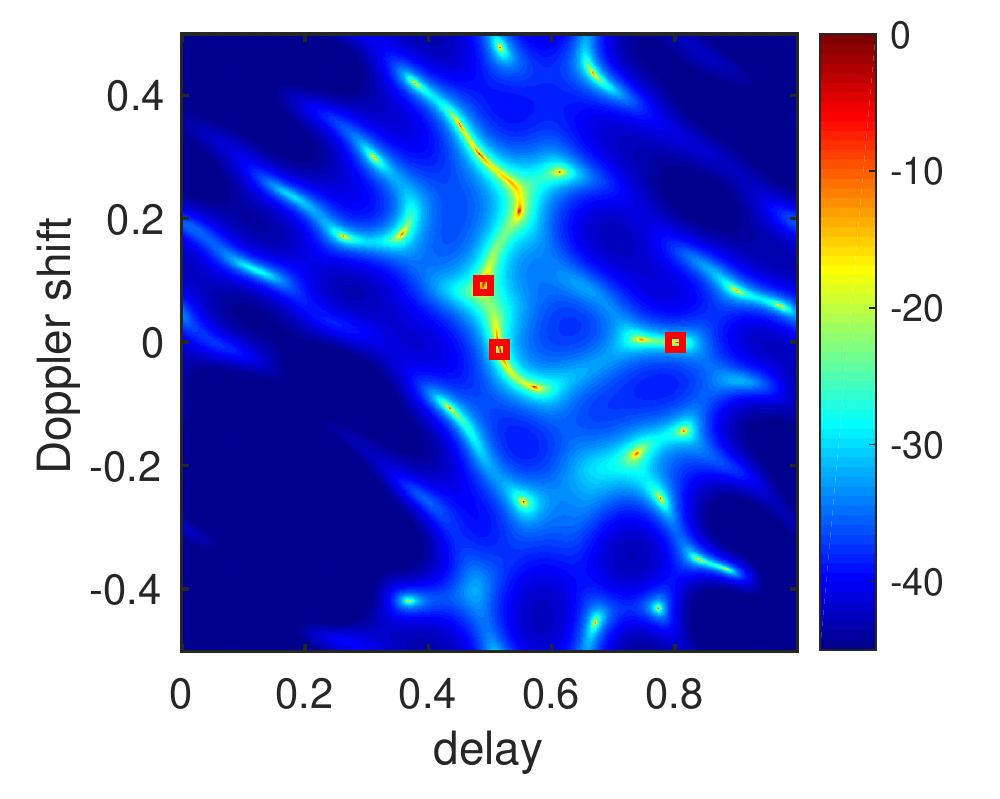}}
	
	\subfloat[][]{\includegraphics[width=1.6in]{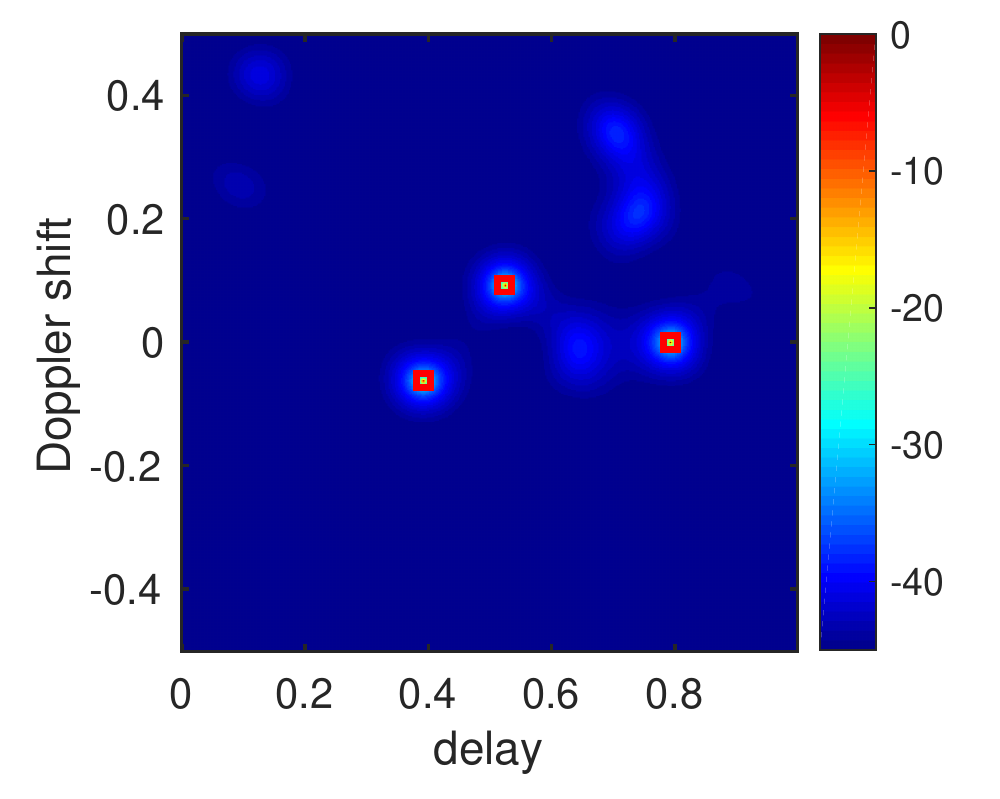}}
	\subfloat[][]{\includegraphics[width=1.6in]{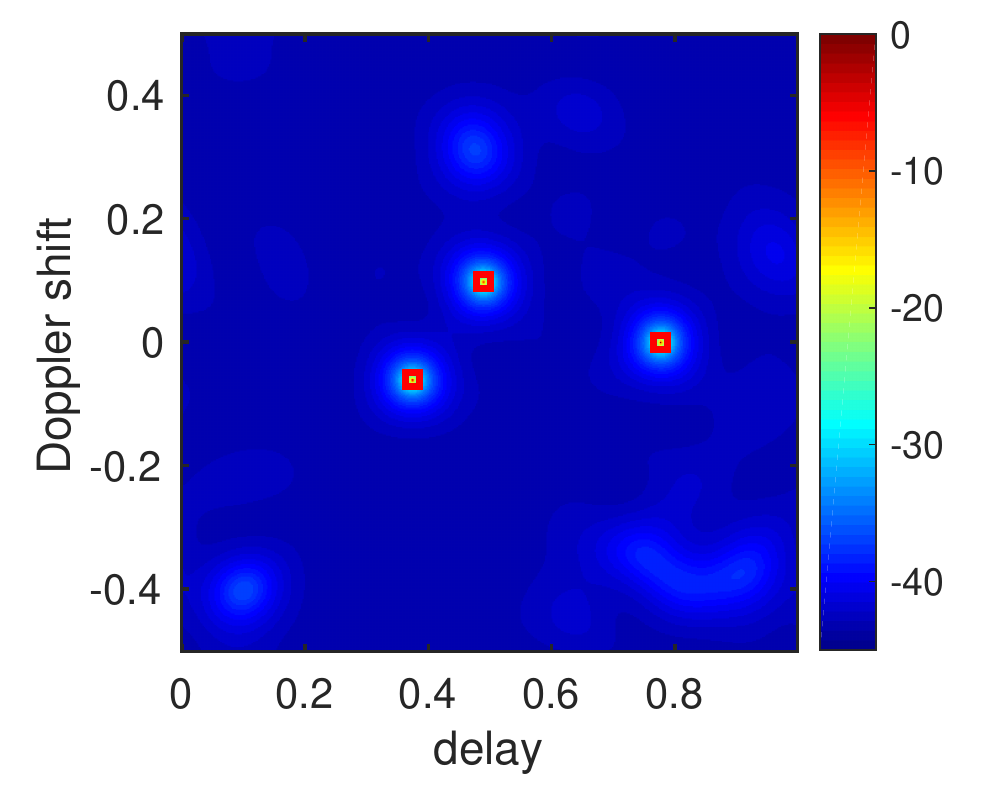}}
	\subfloat[][]{\includegraphics[width=1.6in]{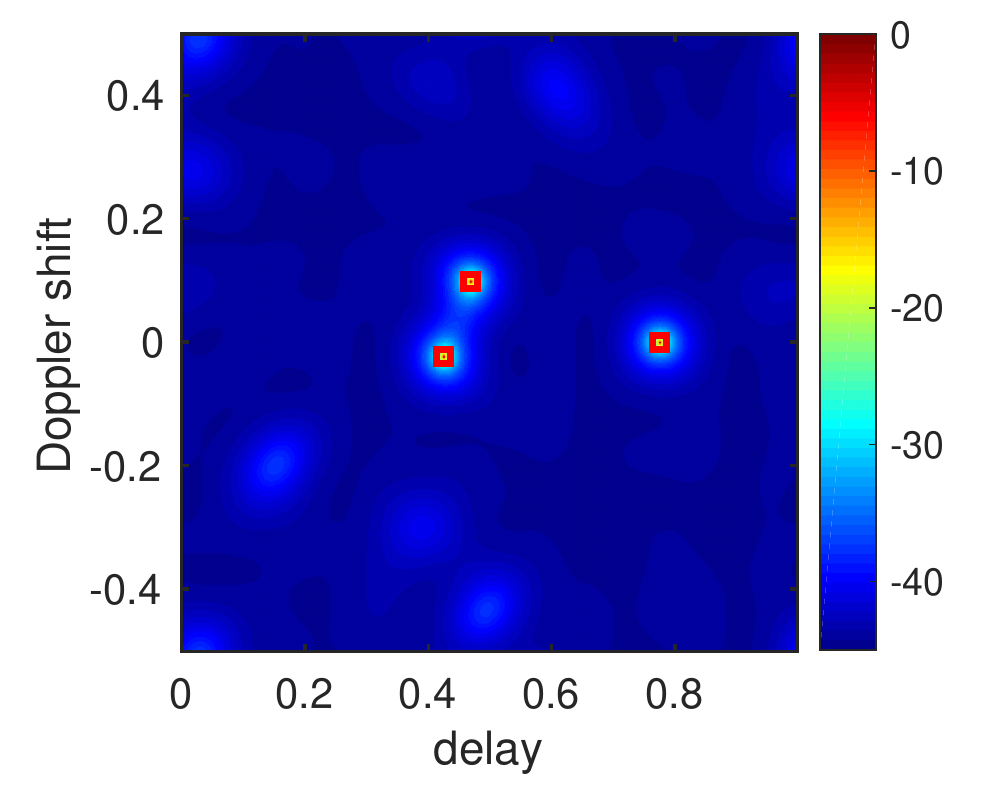}}
	\subfloat[][]{\includegraphics[width=1.6in]{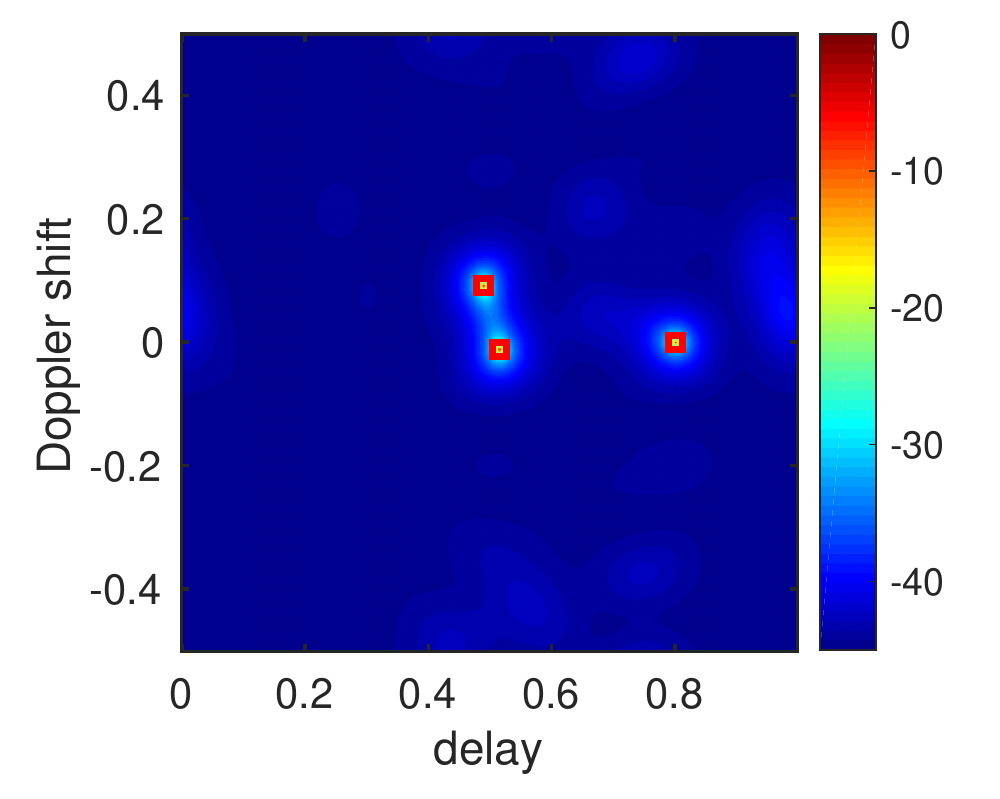}}

	\caption{Delay-Doppler estimation results. (a)-(d) CR estimation results of receivers 1-4 when $\text{BER}=0.01$, respectively; (e)-(h) CR estimation results of receivers 1-4 when $\text{BER}=0.03$, respectively; (i)-(l) CGD estimation results of receivers 1-4 when $\text{BER}=0.03$, respectively. Dark blue represents small values while dark red represents large values. The ground truths of target delays and Doppler shifts are marked by red squares.}
	\label{figure:delay-Doppler-results}
\end{figure*}

The convergence behavior of the proposed CGD method is illustrated next. The SNR in this simulation is set as 20 dB. Fig.~\ref{figure:RMSE-Phi-CGD} shows the $\text{RMSE}_{\Phi}^i$ of the proposed CGD method, the CR method and the CR method when there is no demodulation error (CR-no-error). The running times corresponding to CR and CR-no-error using CVX are 4957.96s and 4141.57s, respectively, while the CGD method only takes 45.51s with 150 iterations. We can see that the performance of the proposed CGD method is close to that of CR-no-error after 150 iterations, while it is significantly faster than the CR method and is suitable for real-time implementation.

%MSE1 = 261.1866 MSE2 = 1.0015  For comparison, we add the $\text{RMSE}_{\Phi}^i$ result of the CR method when the error $\bm e = 0$ is known, i.e., the result obtained by \eqref{eq:SDP-re} when there is no error in $\bm{\hat b}$. 

\begin{figure*}[!htb]
	\centering
		
	\subfloat{\includegraphics[width=2.5in]{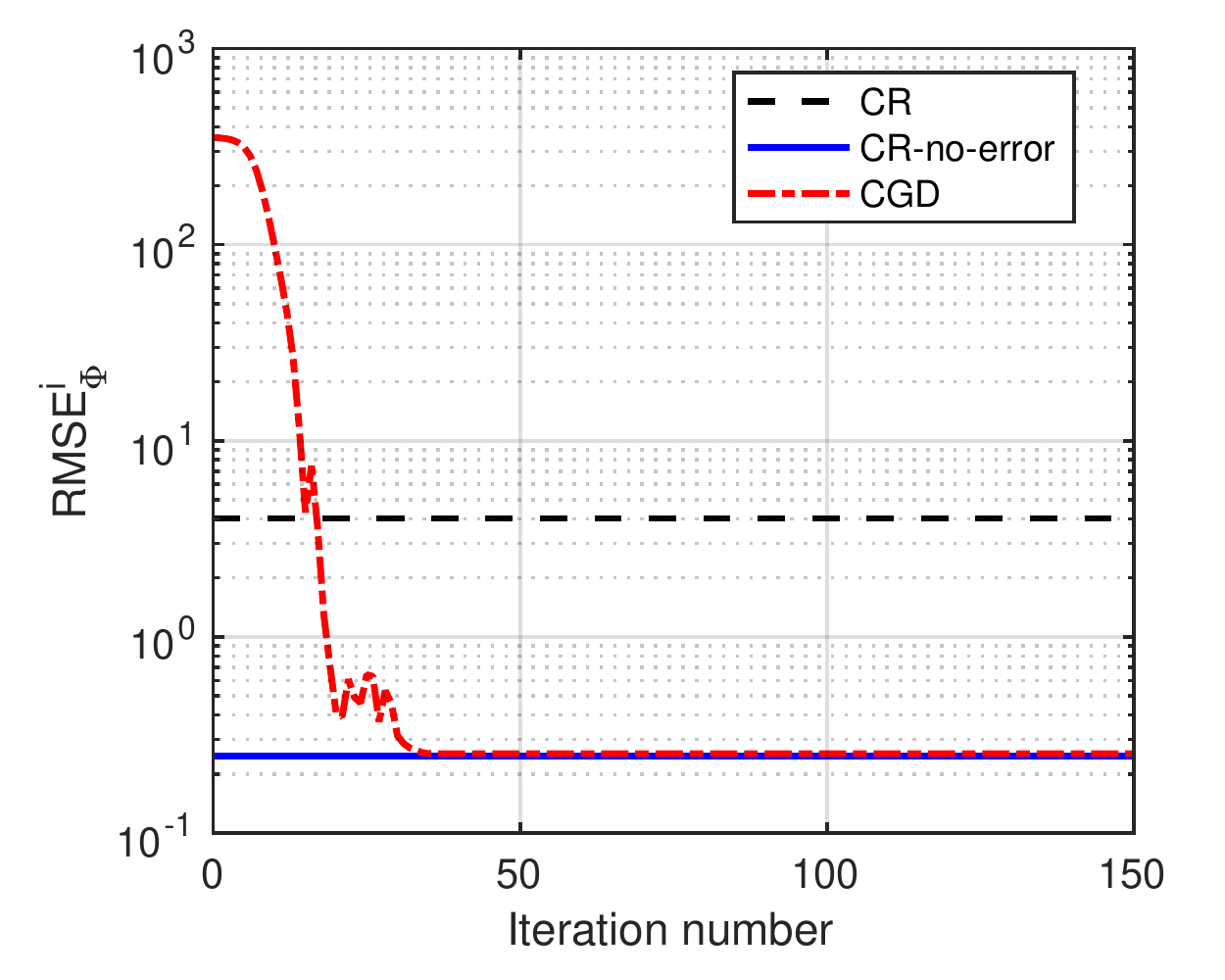}}

	\caption{Convergence behavior of the proposed CGD method and the CR method.}
	\label{figure:RMSE-Phi-CGD}
\end{figure*}

The training times and performances of different BP neural networks are compared in Fig.~\ref{figure:network-result}. Network 1 has one hidden layer with 200 neurons and Network 2 has two hidden layers with 25 neurons in each layer. Comparing the best validation performance for 1000 epochs, we can see that with two hidden layers, we can use fewer neurons than one hidden layer and achieve a better performance. Moreover, 
the training times for 1000 epochs are 876.9s and 198.9s for Network 1 and Network 2, respectively. This is because the number of neurons in Network 2 is significantly smaller than that in Network 1. Hence, a BP neural network with two hidden layers is more suitable for our target localization problem.

\begin{figure*}[!t]
	\centering
		
	\subfloat{\includegraphics[width=2.5in]{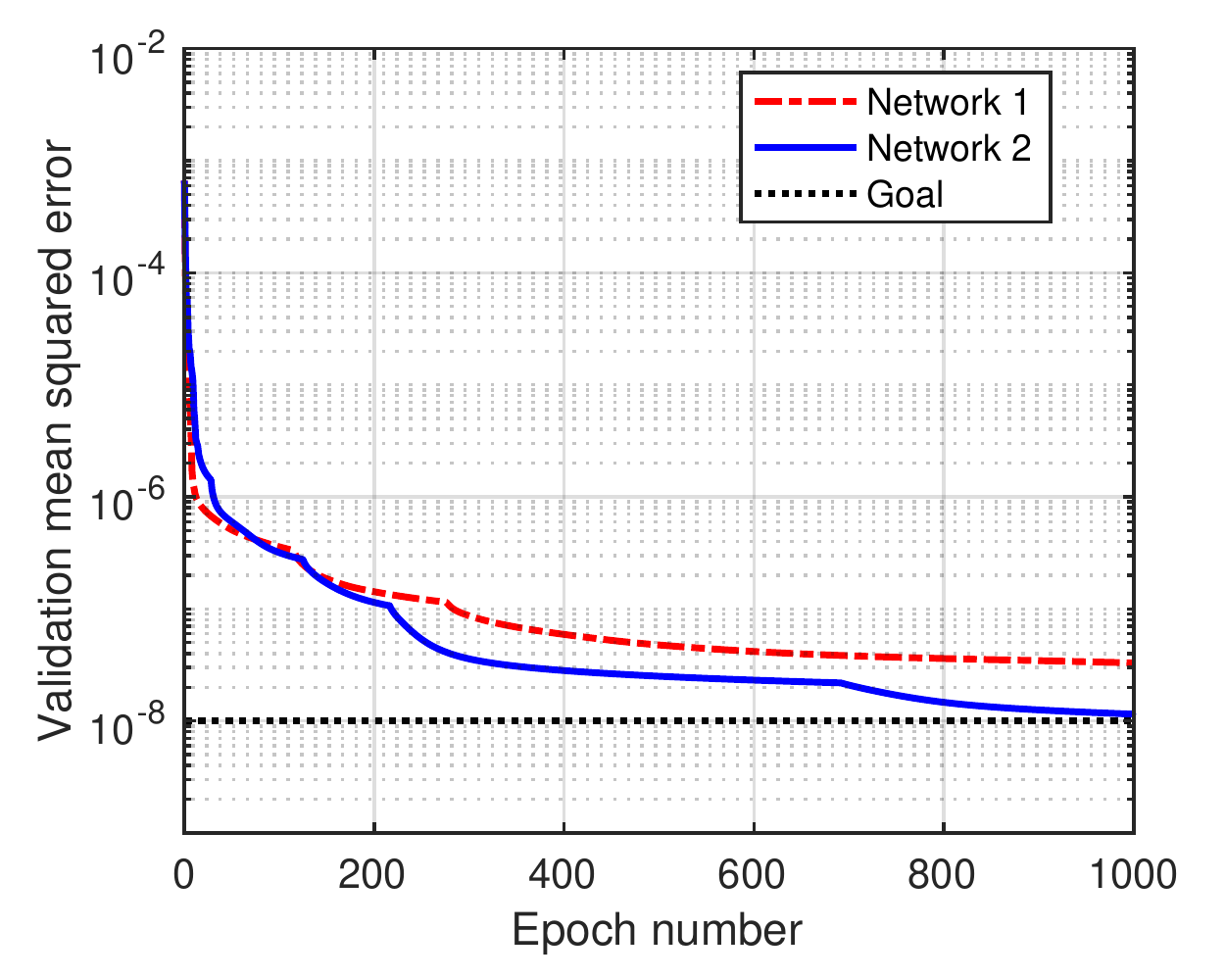}}

	\caption{Training performance of neural networks. Network 1 has one hidden layer with $N_h = 200$ neurons, while Network 2 has two hidden layers with $N_h = 25$ neurons in each layer.}
	\label{figure:network-result}
\end{figure*}

\begin{figure*}[!htb]
	\centering
		
	\subfloat[][]{\includegraphics[width=2.1in]{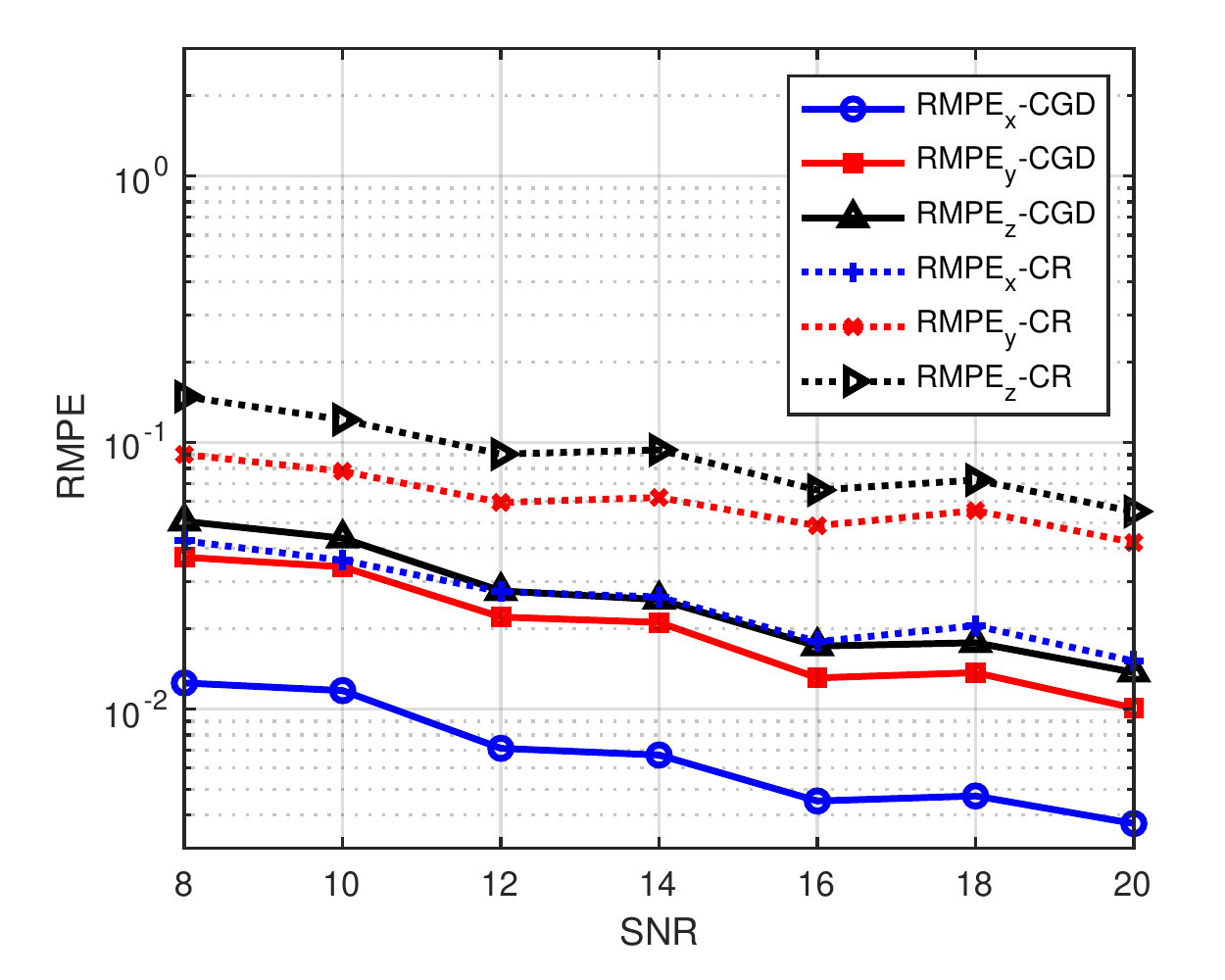}}
	\subfloat[][]{\includegraphics[width=2.1in]{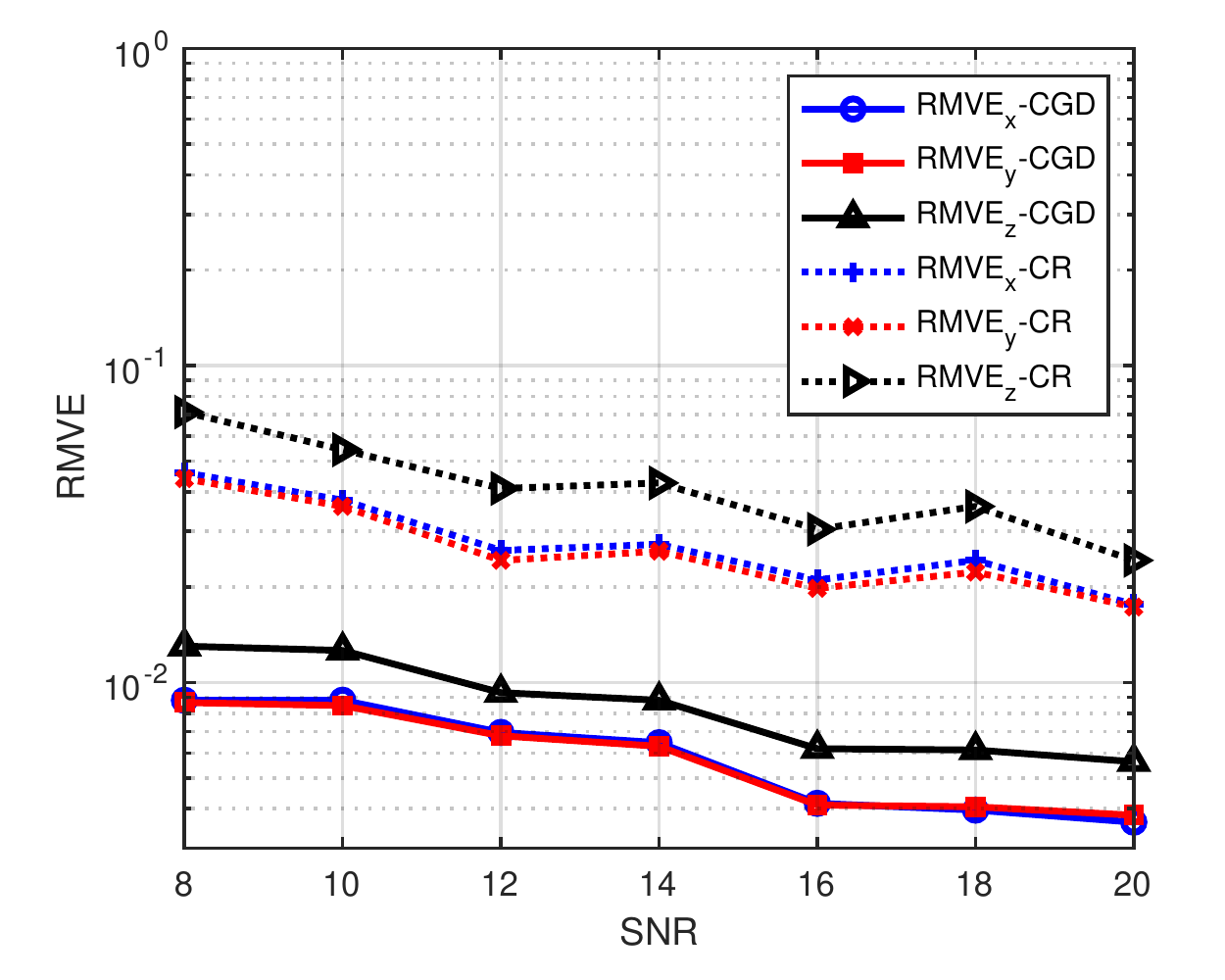}}
	
	\caption{Plots of RMPE and RMVE against SNR. (a) RMPE; (b) RMVE.}
	%\caption{Plots of $\text{RMSE}_{x}$ and $\text{RMSE}_{v}$ under different SNRs. The SNR from bottom left to top right in the curve is 20dB, 14dB and 8dB, respectively.}
	\label{figure:RMSE-SNR}
\end{figure*}

\begin{figure*}[!htb]
	\centering
	
	%\subfloat{\includegraphics[width=3.2in]{RMSEx-RMSEv-BER.pdf}}

	\subfloat[][]{\includegraphics[width=2.1in]{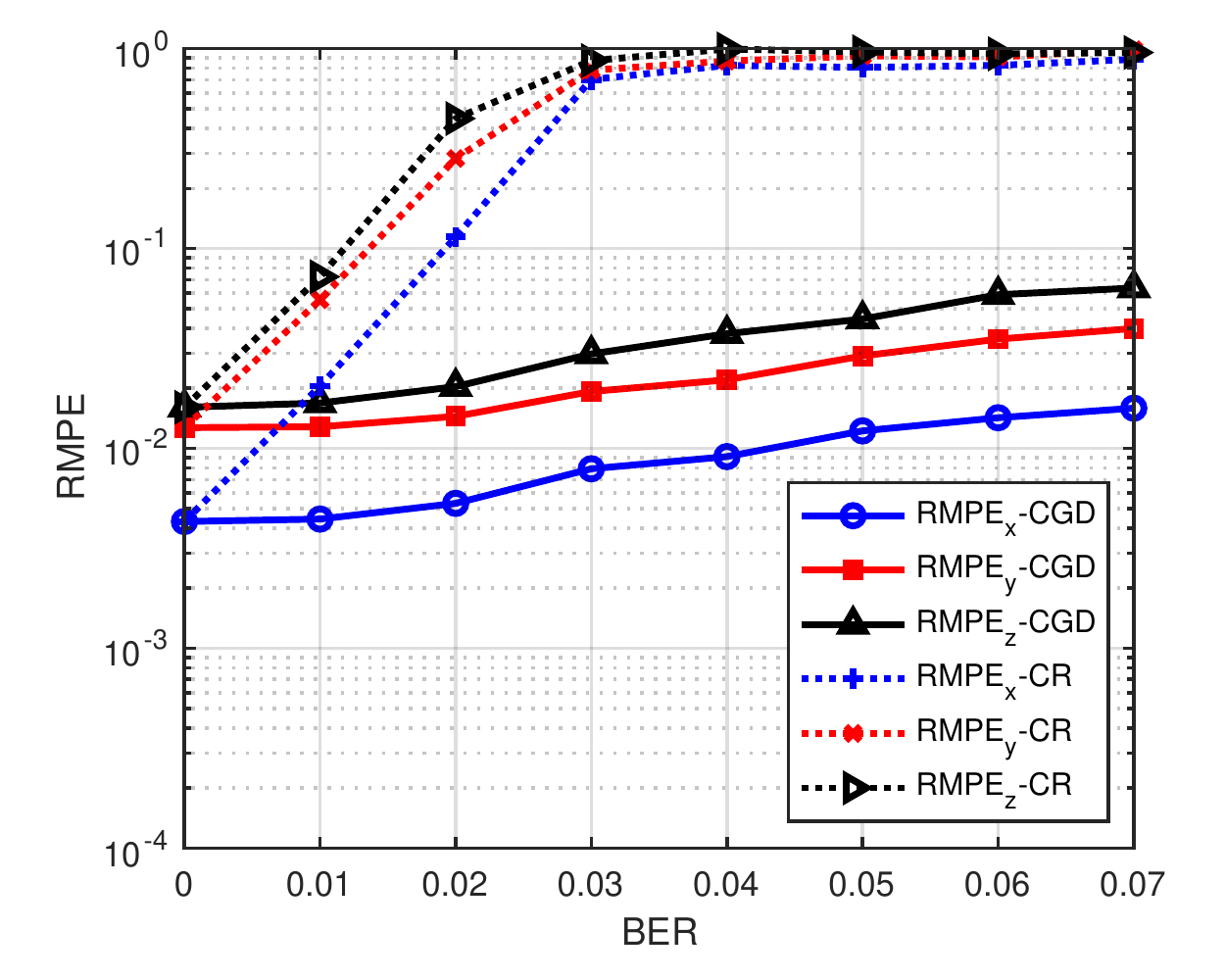}}
	\subfloat[][]{\includegraphics[width=2.1in]{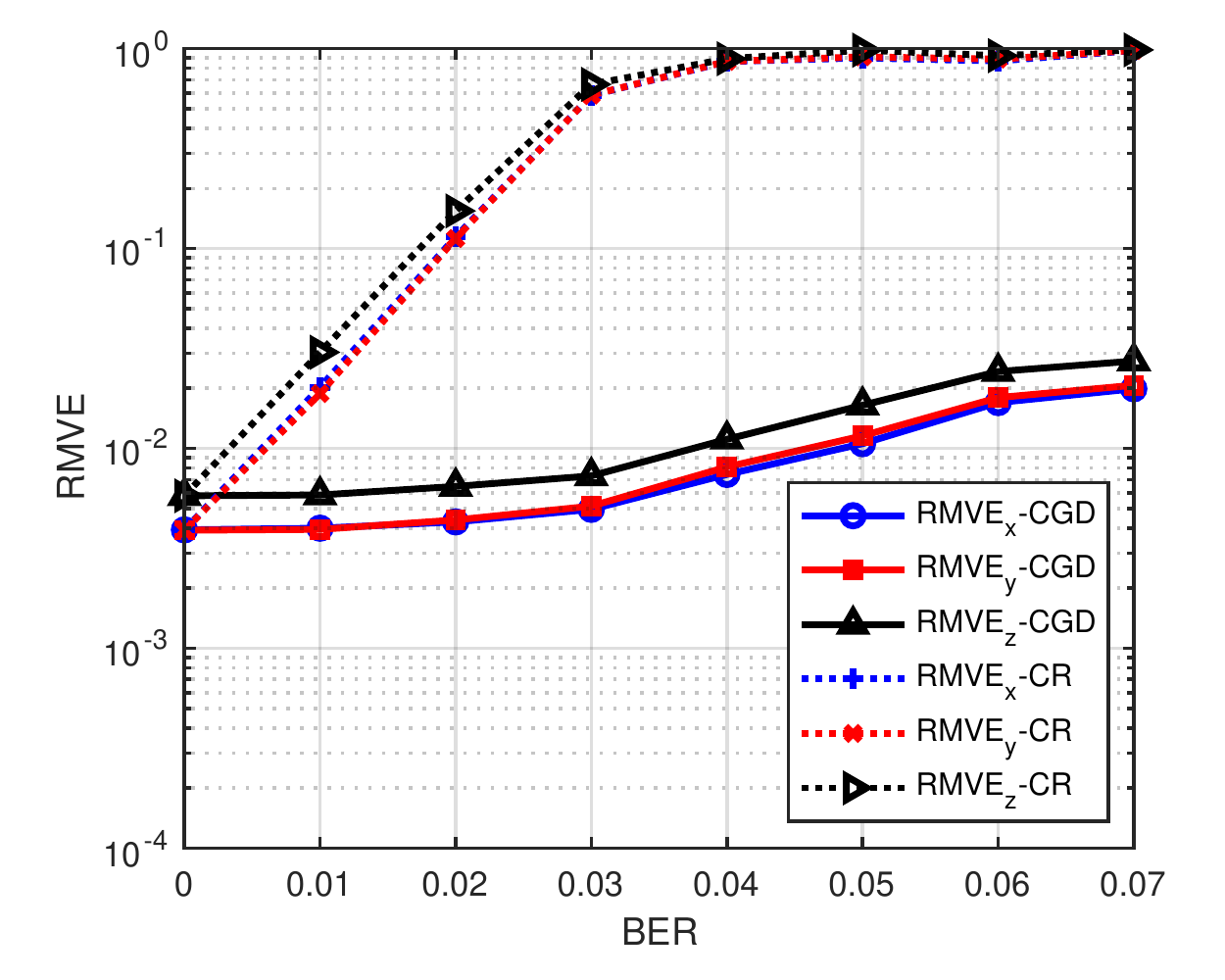}}
	
	\caption{Plots of RMPE and RMVE against BER. (a) RMPE; (b) RMVE.}
	\label{figure:RMSE-BER}
\end{figure*}

Fig.~\ref{figure:RMSE-SNR} shows the RMPE and RMVE of the proposed methods against SNR. The BER is set as 0.01. Since simulation results show the performances of the BP method and the ``solver'' method are very close, we plot only the BP performance. We can see that even for very small BER, the estimation performances of the CR method have a significant degradation compared to the CGD method. Fig.~\ref{figure:RMSE-BER} shows the RMPE and RMVE of the proposed methods against BER. The SNR is set as 15dB. We can see that the CGD method is robust to demodulation errors, and the target positions and velocities can be accurately estimated even for large BER. In contrast, the CR method is no longer effective for large BER since it results in large estimation errors. Note that the RMPEs and RMVEs in Fig.~\ref{figure:RMSE-SNR} and Fig.~\ref{figure:RMSE-BER} along different axes are different, e.g., the position estimation performance along the $x$-axis is best. This is because the receivers and transmitter are placed along the $x$-axis in the simulation and they are well separated (see Fig.~\ref{figure:tracking-results-1}). Moreover, we see that the position estimation performance along the $x$-axis is better than that along the $y$-axis, while the velocity estimation performances along the $x$-axis and the $y$-axis are close. This indicates that the position estimation accuracy and the velocity estimation accuracy are not directly related.

\begin{figure*}[!t]
	\centering
		
	\subfloat{\includegraphics[width=2.1in]{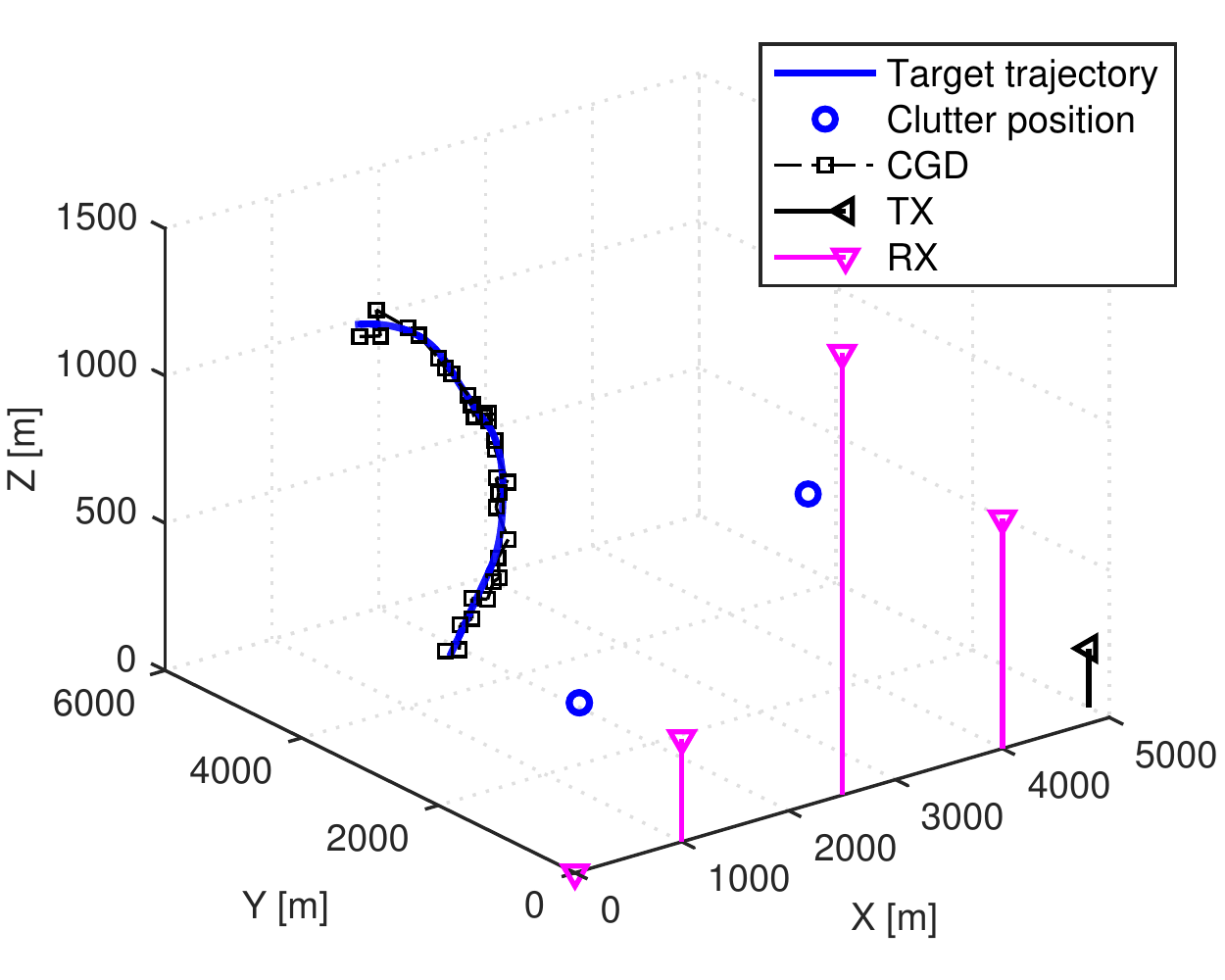}}
	\subfloat{\includegraphics[width=2.1in]{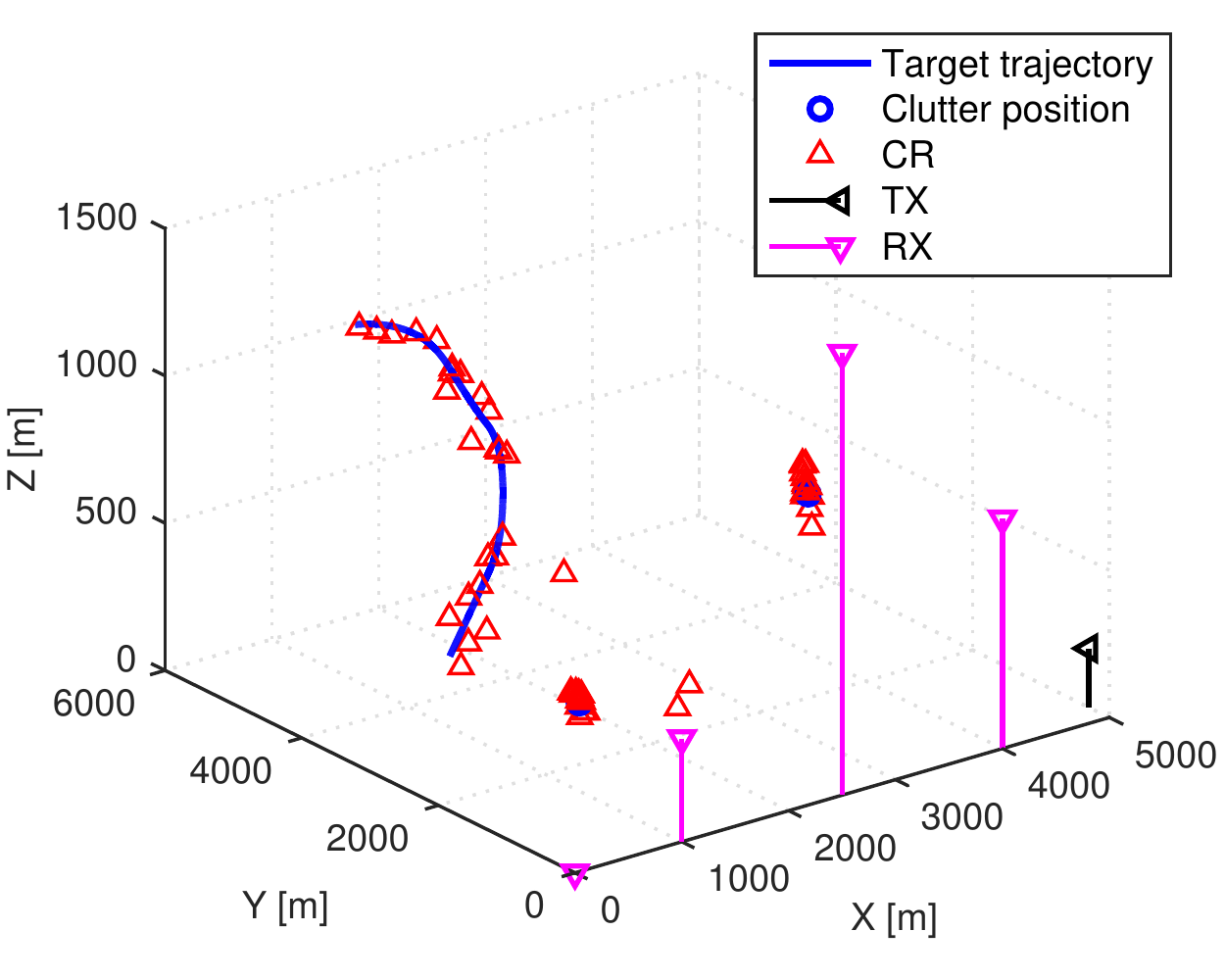}}
	\subfloat{\includegraphics[width=2.1in]{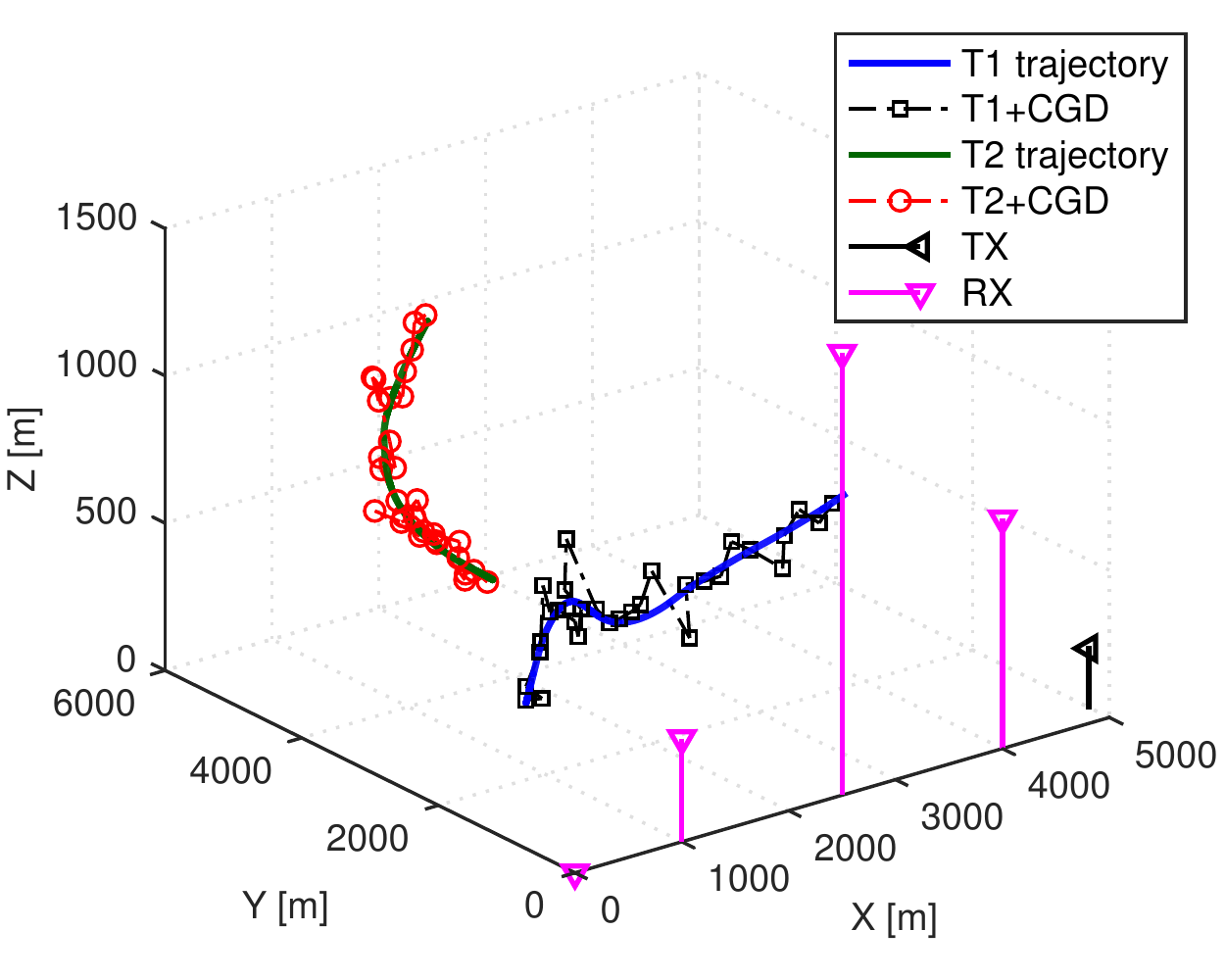}}
	
	\caption{Tracking results. (a) CGD, scenario 1; (b) CR, scenario 1. (c) CGD, scenario 2.}
	\label{figure:tracking-results-1}
\end{figure*}

%\begin{figure*}[!t]
%	\centering
%		
%	\subfloat{\includegraphics[width=2.1in]{Z-1T-X-CGD.pdf}}
%	\subfloat{\includegraphics[width=2.1in]{Z-1T-X-CR.pdf}}
%	%\subfloat{\includegraphics[width=3.2in]{A-1T-X-CGD.eps}}
%	%\subfloat[][]{\includegraphics[width=3.2in]{A-1T-X-CR-V1.eps}}
%	
%	\caption{Tracking results of scenario 1. (a) CGD; (b) CR.}
%	\label{figure:tracking-results-1}
%\end{figure*}
%
%\begin{figure*}[!t]
%	\centering
%		
%	\subfloat{\includegraphics[width=2.1in]{Z-2T-X.pdf}}
%	%\subfloat{\includegraphics[width=3.2in]{A-2T-X.eps}}
%	
%	\caption{Tracking result of scenario 2.}
%	\label{figure:tracking-results-2}
%\end{figure*}

\begin{figure*}[!t]
	\centering
		
	\subfloat[][]{\includegraphics[width=2.1in]{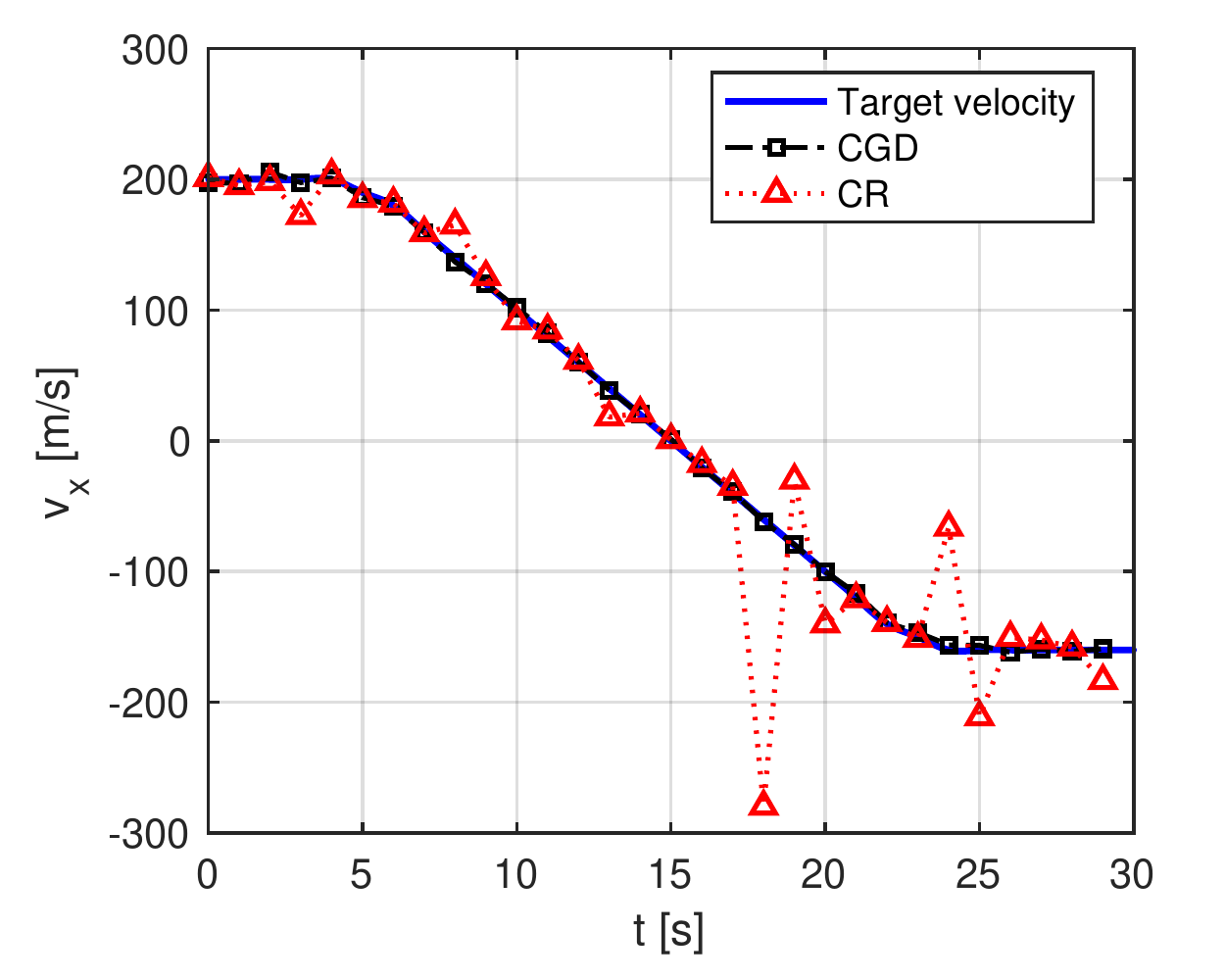}}
	\subfloat[][]{\includegraphics[width=2.1in]{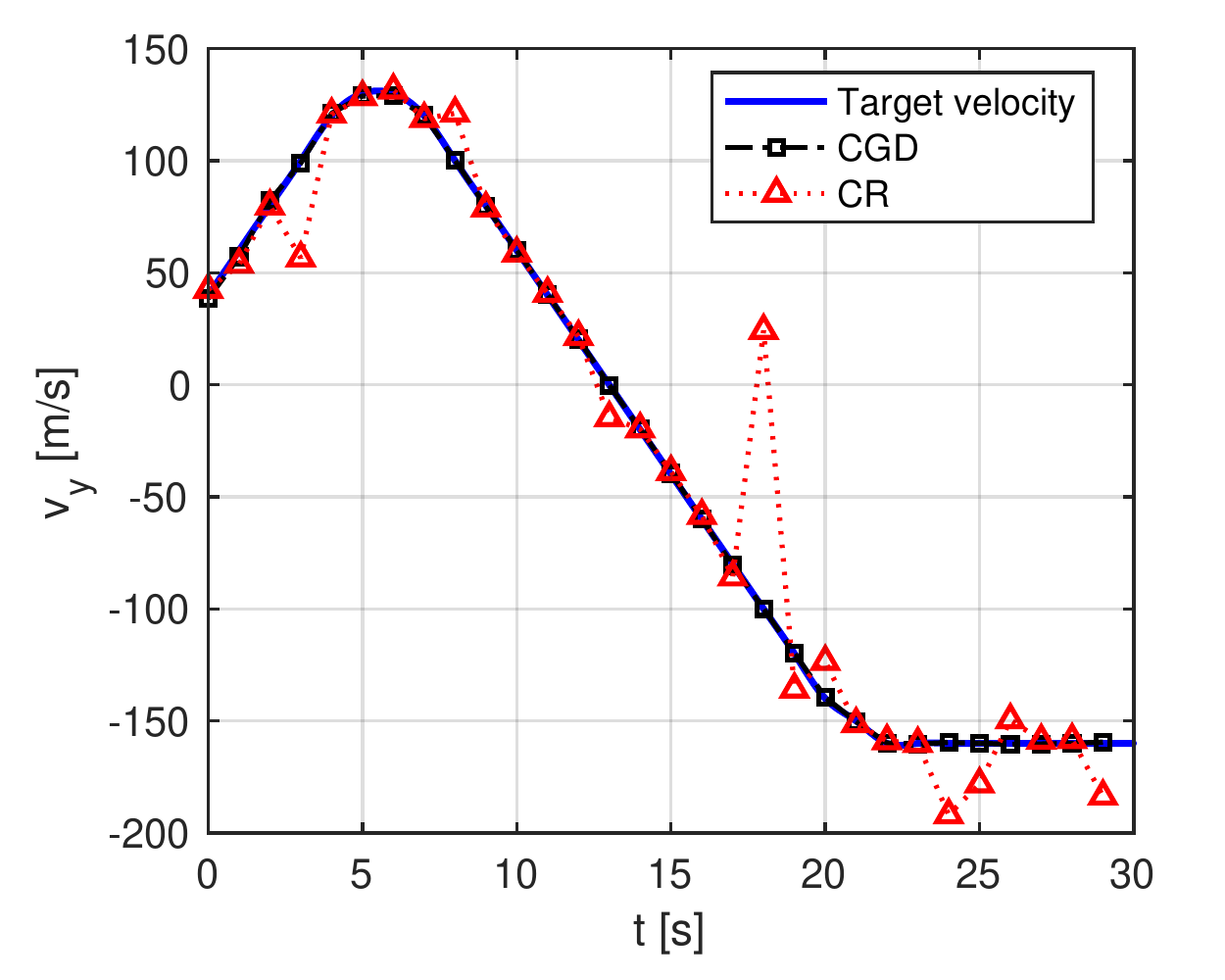}}
	\subfloat[][]{\includegraphics[width=2.1in]{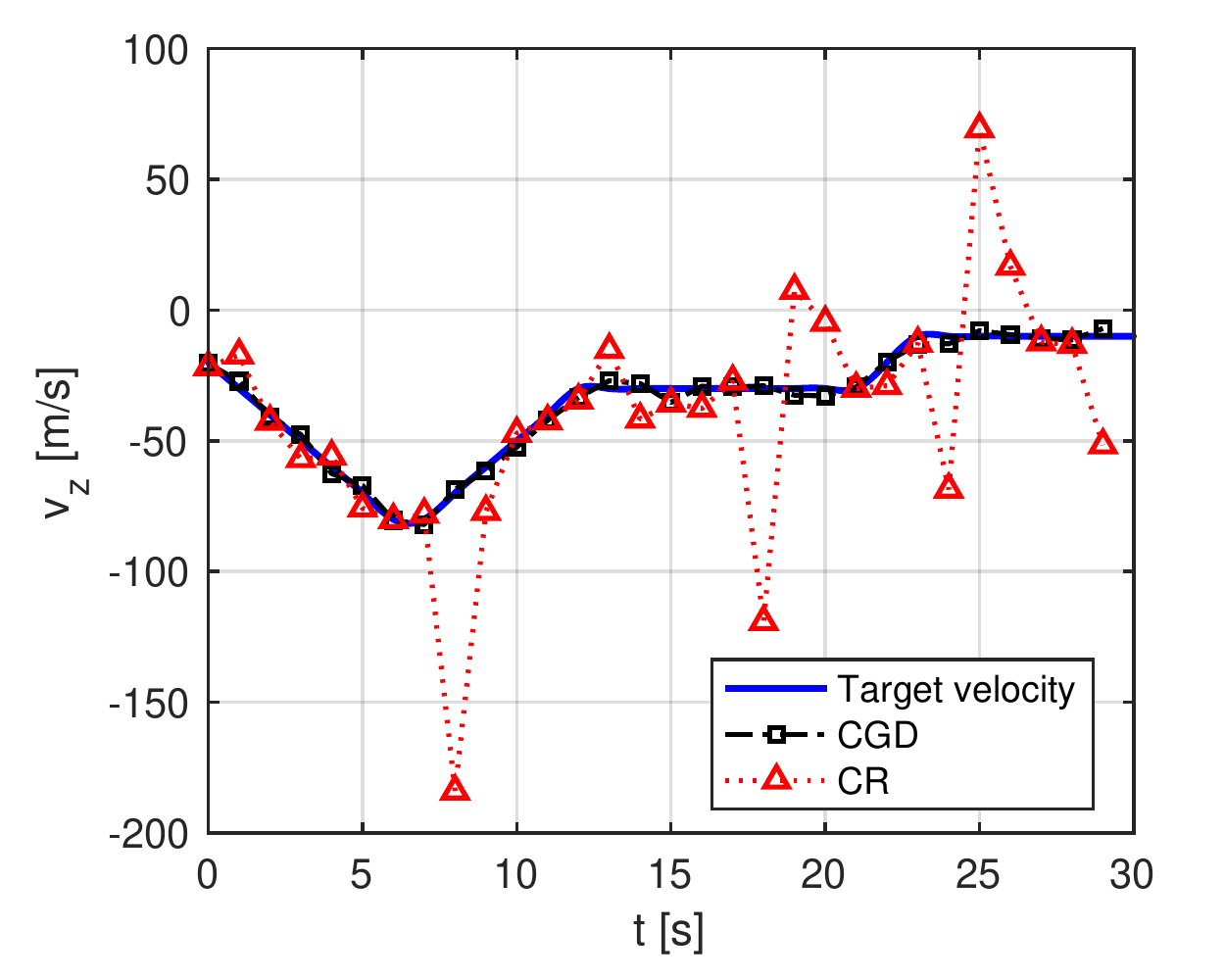}}
	
	\caption{Velocity estimation results of scenario 1. (a) $v_{\ell}^x$; (b) $v_{\ell}^y$; (c) $v_{\ell}^z$.}
	\label{figure:tracking-S1-V}
\end{figure*}

\begin{figure*}[!t]
	\centering
		
	\subfloat[][]{\includegraphics[width=2.1in]{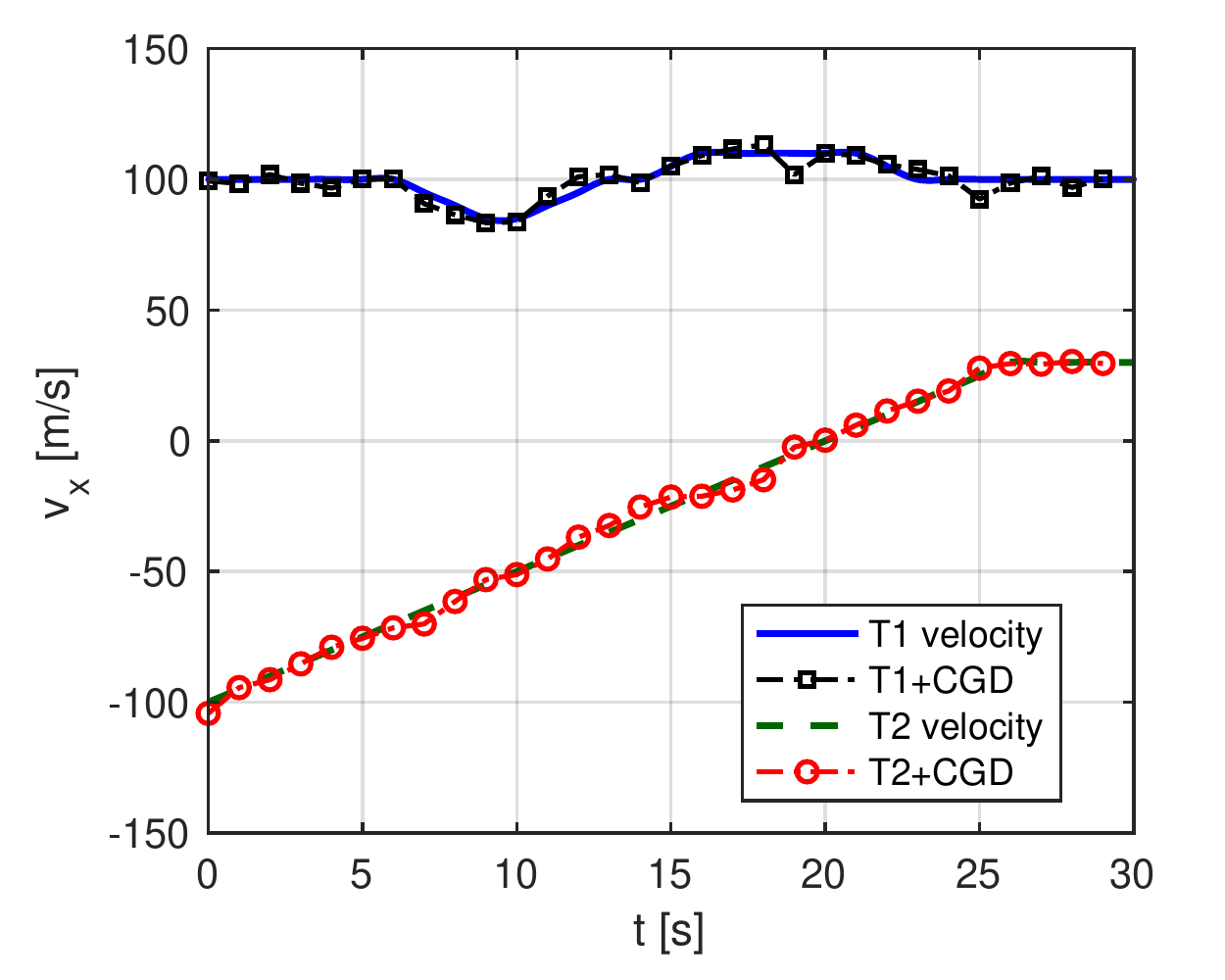}}
	\subfloat[][]{\includegraphics[width=2.1in]{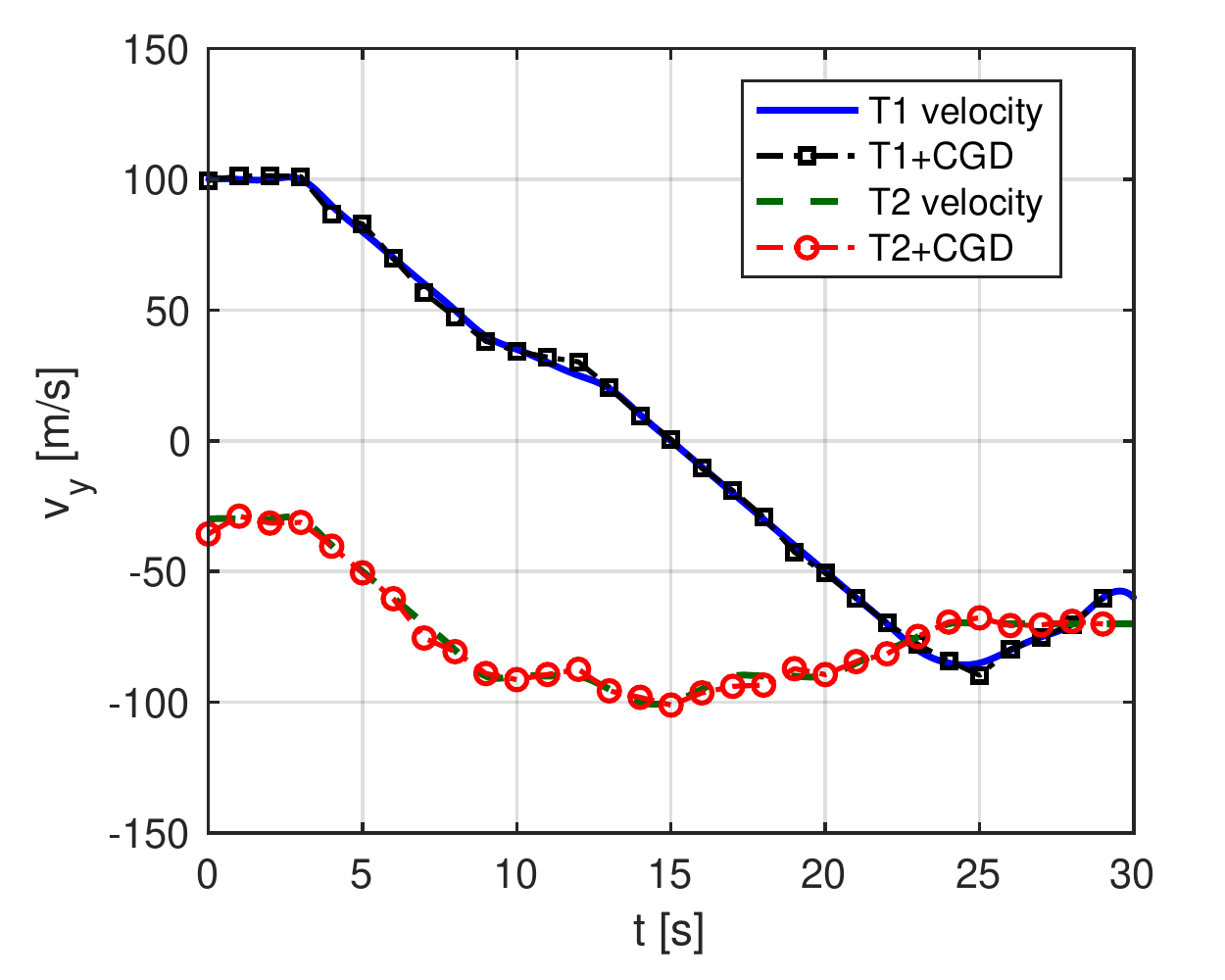}}
	\subfloat[][]{\includegraphics[width=2.1in]{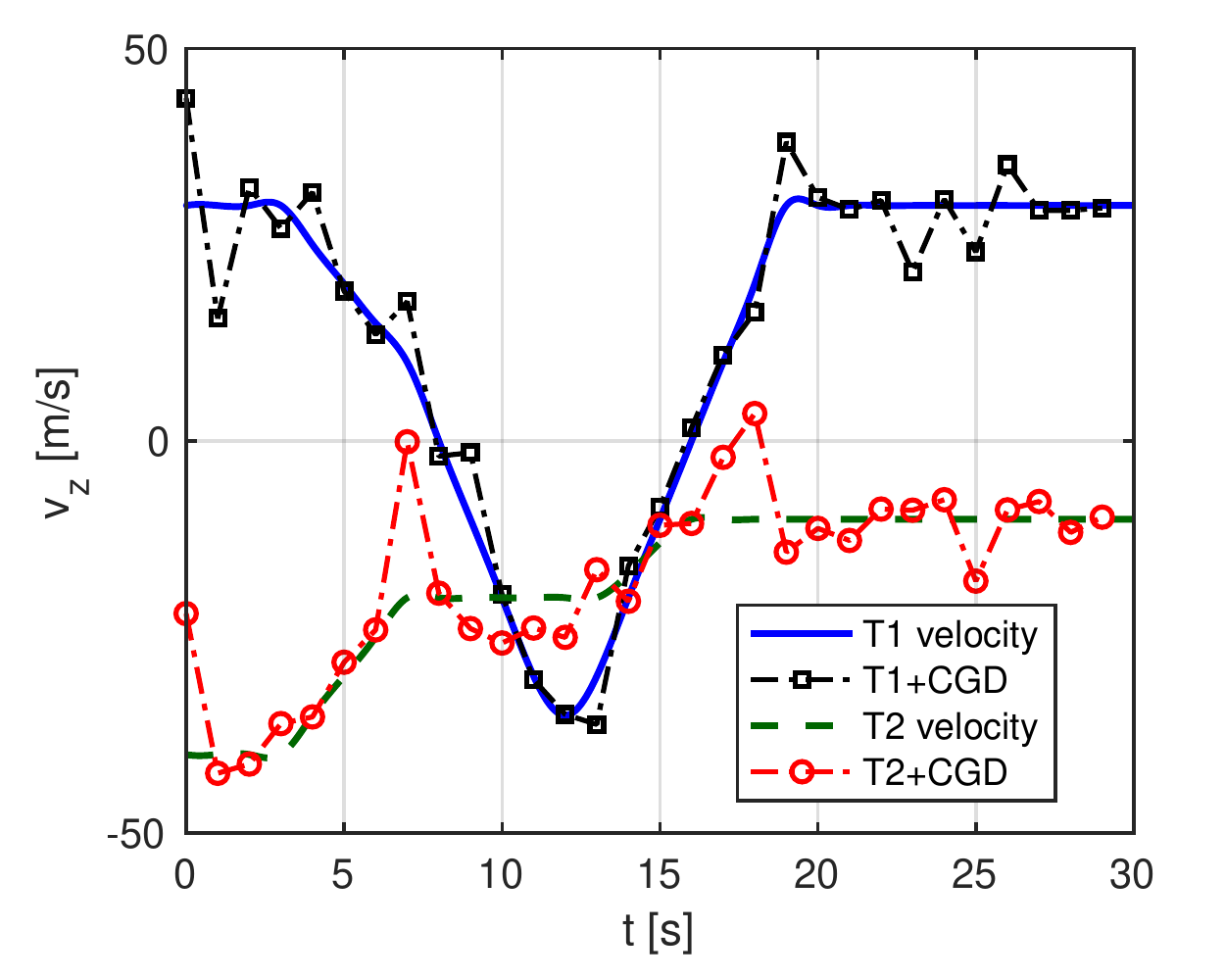}}
	
	\caption{Velocity estimation results of scenario 2. (a) $v_{\ell}^x$; (b) $v_{\ell}^y$; (c) $v_{\ell}^z$.}
	\label{figure:tracking-S2-V}
\end{figure*}

Next, we show the simulation results of tracking maneuvering targets. Figs.~\ref{figure:tracking-results-1}(a)-(b) and Fig.~\ref {figure:tracking-S1-V} show the tracking results and velocity estimation results of scenario 1, respectively. When an estimated velocity of target $\|\bm v_{\ell}\|_2<3$, it is a considered clutter instead of a moving target. We can see that accurate velocity estimates can be obtained by the CGD method, which is very helpful for the identification of clutters. After identifying clutters according to their velocities, the false alarms caused by clutters are all eliminated in Fig.~\ref{figure:tracking-results-1}(a). In contrast, using the CR method, the position and velocity estimation performances are significantly degraded even when the BER is 0.01, due to the inaccuracy in estimating the delays and Doppler frequencies. Hence, we can see that in Fig.~\ref{figure:tracking-results-1}(b), some estimated target positions are far away form the true target positions and some are out of the surveillance area (the number of estimated target positions are less than 30.). In addition, since the velocities cannot be accurately estimated based on the CR method, many clutters are misjudged as targets, and many false alarms appear in Fig.~\ref{figure:tracking-results-1}(b), which affects target recognition. It is noteworthy that under the same BER condition, the target position and velocity estimation performance also fluctuates because the location where the error occurs is random. The estimated velocity is more inaccurate around 18s than that at other times in Fig.~\ref {figure:tracking-S1-V} because the error location at this time may has a greater impact on position and velocity estimation.

Similarly, Fig.~\ref{figure:tracking-results-1}(c) and Fig.~\ref{figure:tracking-S2-V} show the tracking results and velocity estimation results of scenario 2, respectively. Since the CR method cannot obtain the accurate delays and Doppler shifts when $\text{BER}= 0.03$ (see Fig.~\ref{figure:delay-Doppler-results}), we only show the results based on the proposed CGD method. We can see that both the target positions and velocities can be closely tracked. Finally, we summarize the running times under the two tracking scenarios in Table 1. We can find that the proposed CGD method not only performs better than the CR method, but also has moderate computational complexity. Both the BP and ``solver'' methods have good target position estimation performance, but the BP method is significantly faster, which is more suitable for real-time implementation.

 \begin{table}[!htbp]
	\begin{center}
		\caption{Running Times}
		\begin{tabular}{c|cc|cc}
			\hline
			\hline
			Methods & CR/h & CGD/h & ``solver''/s & BP/s   \\
			\hline
			Scenario 1 & 31.6501 & 0.4192 & 57.0914(CR) &  0.0148(CR)  \\
			%\hline
			& & & 59.1917(CGD) & 0.0152(CGD) \\
			\hline
			Scenario 2 & - & 0.4047 & 39.5009(CGD) & 0.0109(CGD) \\
			\hline
			\hline
		\end{tabular}
	\end{center}
\end{table}

\section{Conclusions}

In this paper, we have proposed a two-stage procedure for estimating the positions and velocities of multiple moving targets based on OFDM passive radar. In the first stage, a non-convex optimization problem for estimating the target delays, Doppler shifts and demodulation error is formulated by exploiting sparsities in terms of atomic norm and $\ell_1$-norm. Then the non-convex optimization problem is relaxed to a smooth unconstrained form and solved by the conjugate gradient descent method. In the second stage, two localization methods are considered to determine the target positions based on the delay differences between different receivers. The first method is based on numerically solving a set of nonlinear equations, while the second method is based on the BP neural network. The target velocities can then be obtained using the estimated Doppler shifts and positions. Simulation results show that the proposed methods can provide accurate target position and velocity estimates under different target movement, SNR and BER conditions. The accurate velocity estimate makes it possible to distinguish between moving targets and static clutters, reducing the false alarm in target detection.

%The target tracking performance of the proposed methods is also very good, accurately estimating the target velocities makes it possible to distinguish between targets and clutters, hence the false alarms in the target detection are greatly reduced. Moreover, the proposed conjugate gradient descent method can estimate the target delays and Doppler shifts accurately with moderate computational complexity, both the ``solver'' and BP methods can provide exactly target positions and velocities and the BP method is more computationally efficient.

%In this paper, we have proposed two algorithms for removing the radar interference to facilitate more reliable data demodulation in a communication system overlaid with a radar system. The first one is based on forcing an atomic norm constraint, and estimating the combination of the radar parameters by solving a convex problem under some relaxations. The second algorithm estimates the radar parameters and the communication demodulation errors by two-stage processing. The first stage obtains a local optimum by alternating minimization, and the second stage infers the global optimum in a higher dimensional space by using the estimates of the first stage. The atomic norm and the $\ell_0$-norm are used to exploit the sparsity of the radar signal components and the sparsity of the demodulation error, respectively. Simulation results show that both algorithms provide much better SER performance compared to the conventional on-grid method. Moreover, the proposed 2-AltMin algorithm offers superior performance and is computationally efficient.

\appendix

%\subsection{Armijo Line Search}
%The Armijo line search approach ensures the selected step size is small enough to guarantee a sufficient decrease of the cost function but not too small. In Algorithm 2 we summarize the Armijo line search for calculating $\mu^i$ in \eqref{eq:SUOP}.

%\begin{algorithm}[htbp]\small
%	\label{tab:A3}
%	\caption{Armijo line search}
%	\begin{tabular}{lcl}
%		Input $\bm Z_m^i $, $\bm e^i$, $\bm G_m^i$, $\bm g^i$, ${\varrho} \in (0,1)$, $\bar{\varrho} \in (0,1/2)$, $\bm Z_m^{i-1}$({\em optional}), $\bm e^{i-1}$({\em optional}).\\
%	        1, {\sf{If}} $\bm Z_m^{i-1}$ and $\bm e^{i-1}$ are absent {\sf{then}} $\mu^i=1/(\| {\bm G}_m^i  \|_F^2 + \|\bm g^i\|_2^2)$,\\
%	        2, {\sf{Else}} $\mu^i=1/(\| {\bm G}_m^i  \|_F^2 + \|\bm g^i\|_2^2)$ \\
%	        2, \sf{Repeat} \\
%	        3, \hspace{0.4cm} $\mu^i$ = ${\varrho} \mu^i$ \\
%	        4, \sf{Until} $ \zeta \left((\bm Z_m^{i} + \mu^i \bm G_m^i)(\bm Z_m^{i} + \mu^i {\bm G}_m^i)^H, \bm e^{i} + \mu^i \bm g^i \right) \leq  \zeta(\bm Z_m^i (\bm Z_m^i)^H,\bm e^i) - \bar{\rho} \mu^i  (\| {\bm G}_m^i  \|_F^2 + \|\bm g^i\|_2^2)$. \\
%	        \midrule
%	        \sf{Return} $\mu^i$.
%	\end{tabular}
%\end{algorithm}

\subsection{Proof of Lemma 1}

Suppose $N_bN_d\geq1025$, $\Delta^{\tau,f} \geq \frac{4.76}{N_bN_d}$ and the solution to \eqref{eq:optimization-problem} is $\bm{\hat \Phi}$ where $\bm{\hat\phi}_m = \sum_{\ell=1}^{L}\hat c_{\ell,m} \bm a(\hat \tau_{\ell,m},\hat f_{\ell,m}),~m=1,...,M$.
%\begin{eqnarray}
%\bm{\hat\phi}_m = \sum_{\ell=1}^{L}\hat c_{\ell,m} \bm a(\hat \tau_{\ell,m},\hat f_{\ell,m}),~m=1,...,M.
%\end{eqnarray}
On the one hand, since \eqref{eq:optimization-problem} and \eqref{eq:SDP2} are equivalent, we have $\bm{\hat\Theta}_m,~m=1,...,M$ is the solution to \eqref{eq:SDP2} once
\begin{eqnarray}
\min_{\bm{\hat\Phi}} \sum\limits_{m=1}^{M} \| \bm{\hat\phi}_m \|_{{\cal A}} = \min_{\bm{\hat{U}}_m,\hat\nu_m} \left\{  \frac{1}{2N_bN_d} \sum\limits_{m=1}^{M} {\rm{Tr}}(\bm{\hat{U}}_m) +\frac{1}{2} \sum\limits_{m=1}^{M} \hat\nu_m \right\},
%\sum\limits_{m=1}^{M} \| \bm{\hat\phi}_m \|_{{\cal A}} =  \inf_{\bm{\hat{U}}_m,\hat\nu_m} \left\{ \frac{1}{2N_bN_d} \sum\limits_{m=1}^{M} {\rm{Tr}}(\bm{\hat{U}}_m) +\frac{1}{2} \sum\limits_{m=1}^{M} \hat\nu_m \right\},
\end{eqnarray}
and the constraint ${\mathbb{T}}({\mathbb{P}}(\bm{\hat{U}}_m)) = \bm{\hat{U}}_m,~ \bm{\hat\Theta}_m \succeq 0,~m=1,...,M$ hold, where $\bm{\hat\Theta}_m$, $\bm{\hat{U}}_m$, $\bm{\hat\Phi}$ and $\hat\nu_m$ are related through \eqref{Phi-nu}.
%\begin{eqnarray}
%\sum\limits_{m=1}^{M} \| \bm{\hat\Phi} \|_{{\cal A}} = \mathop {\inf} \left\{ \begin{array}{l}
%\frac{1}{2N_bN_d}{\rm{Tr}}({\mathbb{T}}(\hat{\cal T})) + \frac{1}{2} \sum\limits_{m} \hat\nu_m,\\
%{\rm s.t.} {\mathbb{T}}({\mathbb{P}}(\bm{\hat{U}}_m)) = \bm{\hat{U}}_m,~ \bm{\hat\Theta}_m \succeq 0,~m=1,...,M
%\end{array} \right\},
%\end{eqnarray}

%let $\bm{\hat\phi}_m = \hat c_{\ell,m} \bm a(\hat\tau_{\ell,m},\hat f_{\ell,m}),~m = 1,...,M$, where $\hat c_{\ell,m} = |\hat c_{\ell,m}| e^{i\hat\theta_{\ell,m}}$. Then
On the other hand, let
\begin{eqnarray}
\label{eq:Phi-solution}
\bm{\hat\Theta}_m = \sum_{\ell=1}^{L} |\hat c_{\ell,m}| \left[ {\begin{array}{*{20}{c}}
	{\bm a(\hat \tau_{\ell,m},\hat f_{\ell,m})}\\
	{e^{i\hat\theta_{\ell,m}}}
	\end{array}} \right] 
	\left[ {\begin{array}{*{20}{c}}
	{\bm a(\hat\tau_{\ell,m},\hat f_{\ell,m})}\\
	{e^{i\hat \theta_{\ell,m}}}
	\end{array}} \right]^H = 
 \left[ {\begin{array}{*{20}{c}}
	{\bm{\hat{U}}_m}& \bm{\hat\phi}_m\\
	{\bm{\hat\phi}_m^H}& {\sum\limits_{\ell=1}^{L} |\hat c_{\ell,m}|}
	\end{array}} \right] \succeq 0,~m = 1,...,M,
\end{eqnarray}
where $\hat c_{\ell,m} = |\hat c_{\ell,m}| e^{i\hat\theta_{\ell,m}}$ and $\bm{\hat{U}}_m = \sum\limits_{\ell=1}^{L} |\hat c_{\ell,m}|{\bm a(\hat\tau_{\ell,m},\hat f_{\ell,m})}{\bm a(\hat\tau_{\ell,m},\hat f_{\ell,m})}^H,~m=1,...,M$
%\begin{eqnarray}
%\bm{\hat{U}}_m = \sum\limits_{\ell=1}^{L} |\hat c_{\ell,m}|{\bm a(\hat\tau_{\ell,m},\hat f_{\ell,m})}{\bm a(\hat\tau_{\ell,m},\hat f_{\ell,m})}^H,~m=1,...,M
%\end{eqnarray}
are block Toeplitz matrixs. In this way the constraint ${\mathbb{T}}({\mathbb{P}}(\bm{\hat{U}}_m)) = \bm{\hat{U}}_m,~ \bm{\hat\Theta}_m \succeq 0,~m=1,...,M$ is satisfied. Then, by noting that $\bm{\hat\phi}_m = \sum
\limits_{\ell=1}^{L}\hat c_{\ell,m} \bm a(\hat \tau_{\ell,m},\hat f_{\ell,m})$ is the unique solution that satisfies $\| \bm{\hat\phi}_m \|_{{\cal A}} = \sum\limits_{\ell=1}^{L} |\hat c_{\ell,m}|$ when $N_bN_d\geq1025$ and $\Delta^{\tau,f} \geq \frac{4.76}{N_bN_d}$~\cite{chi2015compressive}, we have
\begin{eqnarray}
\min_{\bm{\hat{U}}_m,\hat\nu_m} \left\{  \frac{1}{2N_bN_d} \sum\limits_{m=1}^{M} {\rm{Tr}}(\bm{\hat{U}}_m) +\frac{1}{2} \sum\limits_{m=1}^{M} \hat\nu_m \right\} =  \sum\limits_{m=1}^{M} \sum\limits_{\ell=1}^{L} |\hat c_{\ell,m}| = \min_{\bm{\hat\Phi}} \sum\limits_{m=1}^{M} \| \bm{\hat\phi}_m \|_{{\cal A}}.
\end{eqnarray}
Hence $\bm{\hat\Theta}_m$ in \eqref{eq:Phi-solution} is the solution to \eqref{eq:SDP2} and each $\bm{\hat\Theta}_m$ is rank-$L$, which completes the proof.

\subsection{Backtracking Line Search}
The backtracking line search approach ensures that the selected step size is small enough to guarantee a sufficient decrease of the cost function but not too small. In Algorithm 2 we summarize the backtracking line search for calculating $\mu^i$ in \eqref{eq:CG-iterations1} and \eqref{eq:CG-iterations2}.

\begin{algorithm}[!tbp]\small
	\label{tab:A3}
	\caption{Backtracking line search}
	\begin{tabular}{lcl}
		Input $\bm Z_m^i $, $\bm e^i$, $\bm G_m^i$, $\bm g^i$, ${\varrho} \in (0,1)$ and $\bar{\varrho} \in (0,1/2)$.\\
	        1, Initialize $\mu^i=1$.\\
	        2, \sf{Repeat} \\
	        3, \hspace{0.4cm} $\mu^i$ = ${\varrho} \mu^i$ \\
	        4, \sf{Until} $\zeta \left( \{(\bm Z_m^{i} + \mu^i \bm G_m^i)(\bm Z_m^{i} + \mu^i {\bm G}_m^i)^H\}_{m=1}^M, \bm e^{i} + \mu^i \bm g^i \right)$\\
	         ~~~~~~~~~~$\leq  \zeta\left(\{\bm Z_m^i (\bm Z_m^i)^H\}_{m=1}^M,\bm e^i\right) - \bar{\rho} \mu^i  (\| {\bm G}_m^i  \|_F^2 + \|\bm g^i\|_2^2)$. \\
	        \midrule
	        \sf{Return} $\mu^i$.
	\end{tabular}
\end{algorithm}

\subsection{Gradient Calculations}

Denote $\bm{\bar{E}} = {\rm diag}(\bm e)$ and $\bm{\bar{B}} = {\rm diag}(\bm{\hat{b}})$. The gradient $\nabla_{\bm e}\zeta$ is given by
\begin{align}
\nabla_{\bm e}\zeta =&~ \nabla_{\bm e}\left\{ \frac{1}{2}\| \bm Y - (\bm{\bar{B}} + \bm{\bar{E}}) \bm\Phi \|_F^2 \right\} + \nabla_{\bm e}[ \eta \phi_{\varpi}(\bm e) ] \nonumber \\
=&~ \frac{1}{2} \nabla_{\bm e}\left \{{\rm Tr}(\bm \Phi^H \bm{\bar{E}}^H \bm{\bar{E}} \bm \Phi )\right\} - \frac{1}{2} \nabla_{\bm e} \left \{{\rm Tr}[\bm \Phi^H \bm{\bar{E}}^H (\bm Y - \bm{\bar{B}} \bm\Phi) ] + {\rm Tr}[(\bm Y - \bm{\bar{B}} \bm\Phi)^H \bm{\bar{E}} \bm \Phi  ] \right\}  + \nabla_{\bm e}[ \eta \phi_{\varpi}(\bm e) ] \nonumber \\
\label{nabla-e-1}
=&~ {\rm diag}\left( \bm{\bar{E}} \bm \Phi \bm \Phi^H - (\bm Y - \bm{\bar{B}} \bm\Phi) \bm \Phi^H \right) + \eta \nabla_{\bm e}[ \phi_{\varpi}(\bm e) ],
\end{align}
where the $n$-th element of $\nabla_{\bm e}[ \phi_{\varpi}(\bm e) ] \in \mathbb{C}^{N_dN_b \times 1}$ is 
\begin{align}
\label{nabla-e-2}
\nabla_{e_n}[ \phi_{\varpi}(\bm e) ] = \frac{\sinh(|e_n|/{\varpi})}{\cosh(|e_n|/{\varpi})} \frac{e_n}{|e_n|}.
\end{align}

Next we calculate $\nabla_{{\bm Z_m}}\zeta$. Note that if we partition ${\mathbb{T}}({{\mathbb{P}}({\bm U_m})})$ into $N_d\times N_d$ blocks, such that the $(b_1,b_2)$-th element of the $(d_1,d_2)$-th block of ${\mathbb{T}}({{\mathbb{P}}({\bm U_m})})$ is denoted as $\bm{\tilde{U}}_{m,d_1,d_2}(b_1,b_2) $. Then we have for $i = -N_d+1,...,N_d-1,~j = -N_b+1,...,N_b-1,~m = 1,...,M$
\begin{align}
%{\mathbb{P}}({\bm U_m})(i,j) =&~ \frac{1}{\beta_{i,j}} \sum_{d1-d2=i}^{b_1-b_2=j} {{\bm{\bar{U}}}_{m,d_1,d_2}}(b_1,b_2),\\
{\bm{\tilde U}_{m,d_1,d_2}}(b_1,b_2) =&~ {\mathbb{P}}({\bm U_m})(i,j),~ d1-d2=i,~ b_1-b_2=j,
\end{align}
where ${\mathbb{P}}({\bm U_m})(i,j)$ is given in \eqref{eq:PUij}. Hence
\begin{align}
\label{eq:calU}
\|{\mathbb{T}}({{\mathbb{P}}({\bm U_m})}) - {\bm U_m} \|_F^2 =&~ \sum_{j = -N_b + 1}^{N_b-1} \sum_{i = -N_d + 1}^{N_d-1}  \left( \sum_{d1-d2=i}^{b_1-b_2=j} \left(  {{\bm{\bar{U}}}_{m,d_1,d_2}}(b_1,b_2) - {\bm{\tilde U}_{m,d_1,d_2}}(b_1,b_2) \right)^2 \right) \nonumber \\
=&~ \sum_{j = -N_b + 1}^{N_b-1} \sum_{i = -N_d + 1}^{N_d-1} \left( \sum_{d1-d2=i}^{b_1-b_2=j} \left({{\bm{\bar{U}}}_{m,d_1,d_2}}(b_1,b_2)\right)^2   -  \frac{1}{\beta_{i,j}}{\left(\sum\limits_{d1-d2=i}^{b_1-b_2=j} {{\bm{\bar{U}}}_{m,d_1,d_2}}(b_1,b_2)\right)^2} \right) \nonumber \\
=&~ \sum_{j = -N_b + 1}^{N_b-1} \sum_{i = -N_d + 1}^{N_d-1} \left( { \bm u_{i,j,m}^H \bm u_{i,j,m} - \frac{1}{\beta_{i,j}} \bm u_{i,j,m}^H \bm i_{\beta_{i,j}} \bm i_{\beta_{i,j}}^H \bm u_{i,j,m} } \right),
\end{align}
where $\bm i_{\beta_{i,j}}$ denote the length-${\beta_{i,j}}$ all-one vector and $\bm u_{i,j,m} \in \mathbb{C}^{{\beta_{i,j}} \times 1}$ is a vector whose $k$-th element is 
\begin{align}
\bm u_{i,j,m}(k) = \left\{
\begin{aligned}
&{\bm{\bar{U}}}_{m,\bar\kappa+i,\bar\kappa} \left(\tilde\kappa+j,\tilde\kappa \right), ~i\geq0, j\geq0,  \\
&{\bm{\bar{U}}}_{m,\bar\kappa+i,\bar\kappa} \left(\tilde\kappa,\tilde\kappa-j \right), ~i\geq0, j<0,  \\
&{\bm{\bar{U}}}_{m,\bar\kappa,\bar\kappa-i} \left(\tilde\kappa+j,\tilde\kappa \right), ~i<0, j\geq0, \\
&{\bm{\bar{U}}}_{m,\bar\kappa,\bar\kappa-i} \left(\tilde\kappa,\tilde\kappa-j \right), ~i<0, j<0,
\end{aligned}
\right.
\end{align}
where $\bar\kappa = \left \lceil \frac{k}{N_b-|j|} \right\rceil$ and $\tilde\kappa = k- (\bar\kappa-1)(N_b-|j|)$ with $\lceil \cdot \rceil$ being the ceiling operator.

%involves all the elements in the set 
%$ \{ {{\bm{\bar{U}}}_{d_1,d_2}}(b_1,b_2,m) \}_{d1-d2=i}^{b_1-b_2=j}$.
%\begin{eqnarray}
%\bm u_{i,j,m} = \left[ {{\bm{\bar{U}}}_{1+i,1}}(1+j,1,m), u_{p_2,q}^{m_1,n},...,u_{p_{N_b-|i|},q}^{m_1,n}, u_{p_1,q}^{m_2,n} , ... ,u_{p_{N_b-|i|},q}^{m_{N_d-|j|},n} \right]^T \in \mathbb{C}^{{\beta_{i,j}} \times 1}\\
%\left[ u_{p_1,q}^{m_1,n},u_{p_2,q}^{m_1,n},...,u_{p_{N_b-|i|},q}^{m_1,n}, u_{p_1,q}^{m_2,n} , ... ,u_{p_{N_b-|i|},q}^{m_{N_d-|j|},n} \right]^T \in \mathbb{C}^{{\beta_{i,j}} \times 1}
%\end{eqnarray}
%with $m_a$ being the $a$-th smallest element among the set $\{m\}_{m-n=j}$ and $p_b$ being the $b$-th smallest element among the set $\{p\}_{p-q=i}$.

Then, by noting \eqref{Phi-nu} and \eqref{eq:calU} we can rewrite \eqref{eq:zeta} as
\begin{align}
&\zeta({\bm \Theta_m},\bm e) = \frac{1}{2}\| \bm Y  - (\bm{\bar{B}} + \bm{\bar{E}}) \bm\Phi \|_F^2  + \frac{\gamma}{2N_bN_d} \sum\limits_{m=1}^{M} {\rm{Tr}}({\bm U_m}) + \frac{\gamma}{2}\sum_{m=1}^M\nu_m + \eta \phi_{\varpi}(\bm e) \nonumber \\
& ~~~~~~~~~~~~~~~+ \frac{\rho}{2} \sum_{m = 1}^{M} \sum_{j = -N_b + 1}^{N_b-1} \sum_{i = -N_d + 1}^{N_d-1} \left( \bm u_{i,j,m}^H \bm u_{i,j,m} - \frac{1}{\beta_{i,j}} \bm u_{i,j,m}^H \bm i_{\beta_{i,j}} \bm i_{\beta_{i,j}}^H \bm u_{i,j,m} \right) \nonumber \\
=&~ \underbrace{ \sum\limits_{m=1}^{M} \left( \frac{\gamma}{2N_bN_d} {\rm{Tr}}({\bm U_m}) +  \frac{\gamma}{2}\nu_m  -  \frac{1}{2}{\rm Tr}\{\bm\phi_m \bm{\bar y}_m^H (\bm{\bar{B}} + \bm{\bar{E}}) \} - \frac{1}{2} \bm\phi_m^H (\bm{\bar{B}} + \bm{\bar{E}})^H \bm{\bar y}_m \right) }_{ \sum\limits_{m=1}^{M} \langle \bm\Upsilon_m,{\bm \Theta_m} \rangle} + \underbrace{ \eta \phi_{\varpi}(\bm e) + \sum_{m=1}^M \frac{1}{2} \bm{\bar y}_m^H \bm{\bar y}_m }_{{\xi}(\bm e)} \nonumber \\
& + \underbrace{ \sum_{m = 1}^{M} \left( { \frac{\rho}{2} \sum_{j = -N_b + 1}^{N_b-1} \sum_{i = -N_d + 1}^{N_d-1} \left( { \bm u_{i,j,m}^H \bm u_{i,j,m} - \frac{1}{\beta_{i,j}} \bm u_{i,j,m}^H \bm i_{\beta_{i,j}} \bm i_{\beta_{i,j}}^H \bm u_{i,j,m} } \right)  + \frac{1}{2}  \bm\phi_m^H (\bm{\bar{B}} + \bm{\bar{E}})^H (\bm{\bar{B}} + \bm{\bar{E}}) \bm\phi_m } \right) }_{ \sum\limits_{m = 1}^{M} \langle {\bm \Theta_m},{\Xi}({\bm \Theta_m}) \rangle/2 },
\end{align}
where ${{\xi}(\bm e)}$ is a function that depends on $\bm e$; and matrices $\bm\Upsilon_m \in\mathbb{C}^{(N_bN_d+1) \times (N_bN_d+1)}$ and ${\Xi}({\bm \Theta_m})\in\mathbb{C}^{(N_bN_d+1) \times (N_bN_d+1)}$ are respectively given by
\begin{align}
\bm\Upsilon_m =&~ \frac{1}{2}\left[ {\begin{array}{*{20}{c}}
	\frac{\gamma}{N_bN_d} \bm I_{N_bN_d} & - (\bm{\bar{B}} + \bm{\bar{E}})^H \bm{\bar y}_m \\
	- \bm{\bar y}_m^H (\bm{\bar{B}} + \bm{\bar{E}}) & \gamma
	\end{array}} \right],~m=1,...,M, \\
{\Xi}({\bm \Theta_m}) =&~ \left[ {\begin{array}{*{20}{c}}
 \Pi({\bm U_m})  &  \frac{1}{2}(\bm{\bar{B}} + \bm{\bar{E}})^H (\bm{\bar{B}} + \bm{\bar{E}})\bm\phi_m \\
\frac{1}{2} \bm\phi_m^H (\bm{\bar{B}} + \bm{\bar{E}})^H (\bm{\bar{B}} + \bm{\bar{E}}) & 0
\end{array}} \right],~m=1,...,M,
\end{align}
where $\bm I_{N_bN_d}$ denotes the $N_bN_d\times N_bN_d$ identify matrix and $\Pi({\bm U_m})$ is given by
\begin{align}
\Pi({\bm U_m}) = \frac{\rho}{2} \sum\limits_{j = -N_b + 1}^{N_b-1} \sum\limits_{i = -N_d + 1}^{N_d-1} {\rm diag}\left( 2( \bm u_{i,j,m} - \frac{1}{\beta_{i,j}} \bm i_{\beta_{i,j}} \bm i_{\beta_{i,j}}^H \bm u_{i,j,m}), i, j \right),~m=1,...,M,
\end{align}
with ${\rm diag}(\bm x,i,j)$ outputing an $N_bN_d\times N_bN_d$ matrix of $N_d\times N_d$ blocks whose $j$-th subdiagnal of the $i$-th subdiagnal block is the input vector $\bm x$, and all other elements are zero.

Since $\langle {\bm \Theta_m},{\Xi}({\bm \Theta_m}) \rangle/2$ is a quadratic form, after some manipulations we can have
\begin{align}
\nabla_{\bm \Theta_m}\left(\sum\limits_{m = 1}^{M} \langle {\bm \Theta_m},{\Xi}({\bm \Theta_m}) \rangle/2 \right)  = \nabla_{\bm \Theta_m}(\langle {\bm \Theta_m},{\Xi}({\bm \Theta_m}) \rangle/2) = {\Xi}({\bm \Theta_m}).
\end{align}
Following the chain rule, we can finally obtain
\begin{align}
\nabla_{{\bm Z_m}}\zeta =&~ 2(\nabla_{{\bm \Theta_m}}\zeta |_{{\bm \Theta_m}= {\bm Z_m}\bm Z_m^H} ){\bm Z_m}  \nonumber \\
=&~ 2\left\{\nabla_{{\bm \Theta_m}}\left(\sum\limits_{m = 1}^{M} \langle \bm\Upsilon_m, {\bm \Theta_m} \rangle \right) + \nabla_{\bm \Theta_m}\left(\sum\limits_{m = 1}^{M} \langle {\bm \Theta_m},{\Xi}({\bm \Theta_m}) \rangle/2 \right) + \nabla_{{\bm \Theta_m}}({{\xi}(\bm e)})\right\} {\bm Z_m}  \nonumber \\
\label{nabla-Z-1}
=&~ 2( \bm\Upsilon_m + {\Xi}({\bm \Theta_m}) ) {\bm Z_m},~m=1,...,M.
\end{align}

%Then, we have for $m=1,...,M$,
%\begin{align}
%&~ \nabla_{{\bm \Theta_m}_{a,b,m}}(\langle {\bm \Theta_m},{\Xi}({\bm \Theta_m}) \rangle/2) \nonumber \\
%=&~ \left\{
%\begin{aligned}
%&\nabla_{{{\bm{\bar{U}}}_{d_1,d_2}}(b_1,b_2,m)} \left [ \frac{\rho}{2} \xi(\bm u_{i,j,m}) \right] = \rho ({{\bm{\bar{U}}}_{d_1,d_2}}(b_1,b_2,m) - \frac{1}{\beta_{i,j}} \bm i_{\beta_{i,j}}^H \bm u_{i,j,m} ),~a,b = 1,...,N_bN_d,~{{\bm{\bar{U}}}_{d_1,d_2}}(b_1,b_2,m) \in \bm u_{i,j,m},  \\
%&\nabla_{ \bm\phi_{a,m}} \left [ \frac{1}{4} \bm\phi_m^H (\bm{\bar{B}} + \bm{\bar{E}})^H (\bm{\bar{B}} + \bm{\bar{E}}) \bm\phi_m \right ] = [ \bm{\tilde{B}}^H \bm{\tilde{B}} \bm\nu ]_a,  a = 1,...,N_bN_d, b = N_bN_d+1,  \\
%&\nabla_{ \bm\phi_{b,m}^H} \left [ \frac{1}{4} {\rm Tr}\left( (\bm{\bar{B}} + \bm{\bar{E}})^H (\bm{\bar{B}} + \bm{\bar{E}}) \bm\phi_m\bm\phi_m^H \right) \right ] = [\bm\nu^H \bm{\tilde{B}}^H \bm{\tilde{B}} ]_b,  b = 1,...,N_bN_d, a = N_bN_d+1,  \\
%& \nabla_{\nu} \left[ \langle \bm \Phi, {\cal Q}(\bm\Phi) \rangle/2 \right] = 0,  a = b = N_bN_d+1.
%\end{aligned}
%\right.
%\end{align}

%with $\bm\Phi_{a,b}$ denotes the $(a,b)$-th element of $\bm\Phi$. 
%
%Then we have %Following the chain rule, calculating $\nabla_{{\bm Z_m}}\zeta$ can be converted to the problem of calculating $\nabla_{\bm \Phi}\zeta$, i.e.,

\bibliographystyle{IEEEtran}
\bibliography{database} 
\end{document}